%% file: main.tex
\documentclass[conference]{IEEEtran}
\usepackage{amsmath}
\usepackage{bm}
\usepackage{amssymb}
\usepackage{graphicx}
\usepackage{color, colortbl} 
\usepackage[dvipsnames,svgnames,x11names]{xcolor}
\usepackage{algpseudocode,algorithm}
\usepackage{hyperref}
\usepackage{enumerate}
\usepackage[caption=false,font=footnotesize]{subfig}
\usepackage{cite}
\usepackage{url,comment}
\usepackage[normalem]{ulem}
\usepackage[shortlabels]{enumitem}
\usepackage{amsthm}
\usepackage{pgfplots}
\usepackage{tikz}
\usepackage{mathtools}
\usepackage{balance}
\newcommand{\appropto}{\mathrel{\vcenter{
  \offinterlineskip\halign{\hfil$##$\cr
    \propto\cr\noalign{\kern2pt}\sim\cr\noalign{\kern-2pt}}}}}

\newcommand{\WW}{\bm{W}}
\newcommand{\DD}{\bm{D}(\bm{p})}

\newcommand{\EE}{\bm{E}}
\newcommand{\bb}{\bm{c}}
\newcommand{\aaa}{\bm{a}(\bm{p})} 
 
\newcommand{\hht}{\text{H}}
\newcommand{\ttt}{\top}

\newcommand{\projnull}[1]{\boldsymbol{\Pi}^{\perp}_{#1}}

\bstctlcite{IEEEexample:BSTcontrol}

\pgfplotsset{compat=1.16}

\begin{document}
\bstctlcite{IEEEexample:BSTcontrol}
%\title{Near-field Localization with a\\ Single-antenna RIS Lens}
\title{Near-field Localization with a Reconfigurable Intelligent Surface Acting as Lens}
\author{Zohair Abu-Shaban\IEEEauthorrefmark{1}\IEEEauthorrefmark{2}, Kamran Keykhosravi\IEEEauthorrefmark{2}, Musa Furkan Keskin\IEEEauthorrefmark{2},\\ George C. Alexandropoulos\IEEEauthorrefmark{3}, Gonzalo Seco-Granados\IEEEauthorrefmark{4}, Henk Wymeersch\IEEEauthorrefmark{2}\\
\IEEEauthorrefmark{1}Cohda Wireless Pty Ltd., Australia, \\
\IEEEauthorrefmark{2}Department of Electrical Engineering, Chalmers University of Technology, Sweden \\
\IEEEauthorrefmark{3}Department of Informatics and Telecommunications,
National and Kapodistrian University of Athens, Greece\\
\IEEEauthorrefmark{4}Department of Telecommunications and Systems Engineering, Universitat Autonoma de Barcelona, Spain
}

\maketitle
\begin{abstract}
    Exploiting wavefront curvature enables localization with limited infrastructure and hardware complexity. With the introduction of reconfigurable intelligent surfaces (RISs), new opportunities arise, in particular when the RIS is functioning as a lens receiver. 
    We investigate the localization of a transmitter using a RIS-based lens in close proximity to a single receive antenna element attached to reception radio frequency chain. We perform a Fisher information analysis, evaluate the impact of different lens configurations, and propose a two-stage localization algorithm.  Our results indicate that positional beamforming can lead to better performance when \textit{a priori} location information is available, while random beamforming is preferred when \textit{a priori} information is lacking. Our simulation results for a moderate size lens operating at $28$ GHz showcased that decimeter-level accuracy can be attained within $3$ meters to the lens. 
\end{abstract}

\section{Introduction}
Radio localization and sensing is gaining increased importance in fifth generation (5G) wireless communications systems enabling various commercial applications, such as personal navigation, indoor localization, radar sensing, and robot localization \cite{Positioning_5G_NR}. The combination of increased communication bandwidth and larger antenna arrays in 5G has led to improvements in localization accuracy, rendering efficient localization from a single base station possible \cite{Shahmansoori18TWC}. This trend is continuing in beyond 5G research, where extremely large bandwidths at carrier frequencies up to 0.1 THz are being explored together with transceiver architectures based on extremely massive electromagnetically (EM) excited elements \cite{Samsung} (e.g., conventional dipoles and metamaterials), intended for combating path loss due to small element apertures. The latter factors enable a myriad of new opportunities for radio localization and sensing \cite{TR2019,bourdoux20206g}. 

Reconfigurable intelligent surfaces (RISs) \cite{huang2019reconfigurable}, comprising large numbers of EM excited elements with dynamically tunable phase and/or amplitude, have been recently considered as a candidate technology for sixth generation (6G) wireless communication \cite{Positioning_5G_NR}. An RIS can operate as a smart reflector beyond Snell's law \cite{Marco2019} or as a lens with nearly a continuous phase profile \cite{zeng2014electromagnetic, hu2018beyond,huang2019holographic}.
%\footnote{Different terminologies can be found in the literature, including intelligent reconfigurable/reflective surface (IRS) and large intelligent surface (LIS), where the latter generally refers to a lens with a continuous phase profile.}, 
For the reflector operation mode, currently the most common via nearly passive hardware, RISs are deployed to enable high signal-to-noise ratios (SNRs) in the presence of obstructed line-of-sight. To achieve this goal, sophisticated signal processing is used for both channel estimation and beamforming optimization \cite{Zhang_RIS_CE_2019}. On the other hand, RIS lenses can provide a good trade-off between hardware and signal processing complexity. Due to the usually large size of RISs, the corresponding channel models are distinct from standard multiple input multiple output (MIMO) models, as the following two common assumptions do not hold \cite{bjornson2020power}: (\textit{i}) constant signal power across the surface; and (\textit{ii}) the presence of a planar wavefront. Hence, care must be taken when optimizing and evaluating the performance of communication and localization systems including RISs.  

In the context of localization, RISs have been gaining attention in the past few years and an overview of the main challenges and opportunities for RIS reflectors can be found in \cite{wymeersch2019radio}. The position (PEB) and orientation error bounds (OEB) of a MIMO setup
were evaluated in \cite{elzanaty2020reconfigurable,he2020large} which showed that an RIS can significantly enhance localization performance, provided that RIS phase profiles are designed appropriately. The problem was tackled in \cite{he2020adaptive}, which proposed a hierarchical RIS phase codebook. In \cite{zhang2020metaradar}, a RIS-aided multi-user localization protocol was proposed, based on signal strength measurements. In contrast, pure time-delay measurements were considered in \cite{wymeersch2020beyond}, where RIS phase optimization and RIS selection were evaluated targeting PEB optimization. RIS reflectors have also been applied to sensing, e.g., for posture recognition \cite{hu2020reconfigurable}. 
RIS lens localization has been treated in \cite{hu2018beyond,alegria2019cramer,yang20193,guidi2019radio}. In \cite{hu2018beyond}, the localization Cram\'{e}r-Rao bound for a continuous RIS was computed assuming curvature of the wavefront, revealing the impact of the RIS size, as well as RIS-induced impairments and the effect of different RIS deployments. RIS quantization effects on the localization performance were studied in \cite{alegria2019cramer}. In \cite{guidi2019radio}, the following three different architectures were compared: a RIS lens with a signal antenna, a non-reconfigurable lens with multiple antennas, and a standard planar array. The PEB was derived for all three cases, indicating that accurate localization is possible with a large RIS lens with low hardware complexity. There is also related work, such as \cite{yang20193,shaikh2019radio}, on non-reconfigurable lens localization under the planar wavefront assumption.

In this paper, we study 3-dimensional (3D) localization using the low complexity RIS lens architecture of \cite{guidi2019radio}, which consists of reconfigurable discrete RIS lenses and a single antenna attached to a receive radio frequency (RF) chain, Considering a near-field channel model, we present the Fisher information analysis for the problem at hand and show that PEB depends on the RIS phase profiles. In addition, a low complexity location estimator is designed whose performance is evaluated over a realistic channel model at 28 GHz.

\emph{Notations:} Vectors are denoted in bold letters, whereas matrices in bold capital letters, $\bm{X}_{i:j,k:l}$ returns a matrix comprising rows $i$ through $j$ and columns $k$ through $l$ from $\bm{X}$. Operator $\circ$ is the point-wise product of vectors, $\mathbb{E}\{\cdot\}$ denotes expectation, $\cdot^\dagger$ the matrix pseudo-inverse, and $\jmath=\sqrt{-1}$. $\mathbf{1}_N$ and $\mathbf{0}_N$ represent the all-ones and all-zeros column vectors, respectively, of size $N$. The probability density function (pdf) of a random vector $\bm{x}$ is denoted by $p(\bm{x})$.

\section{System and Channel Models}
\begin{figure}
    \centering
    \includegraphics[width=0.85\columnwidth]{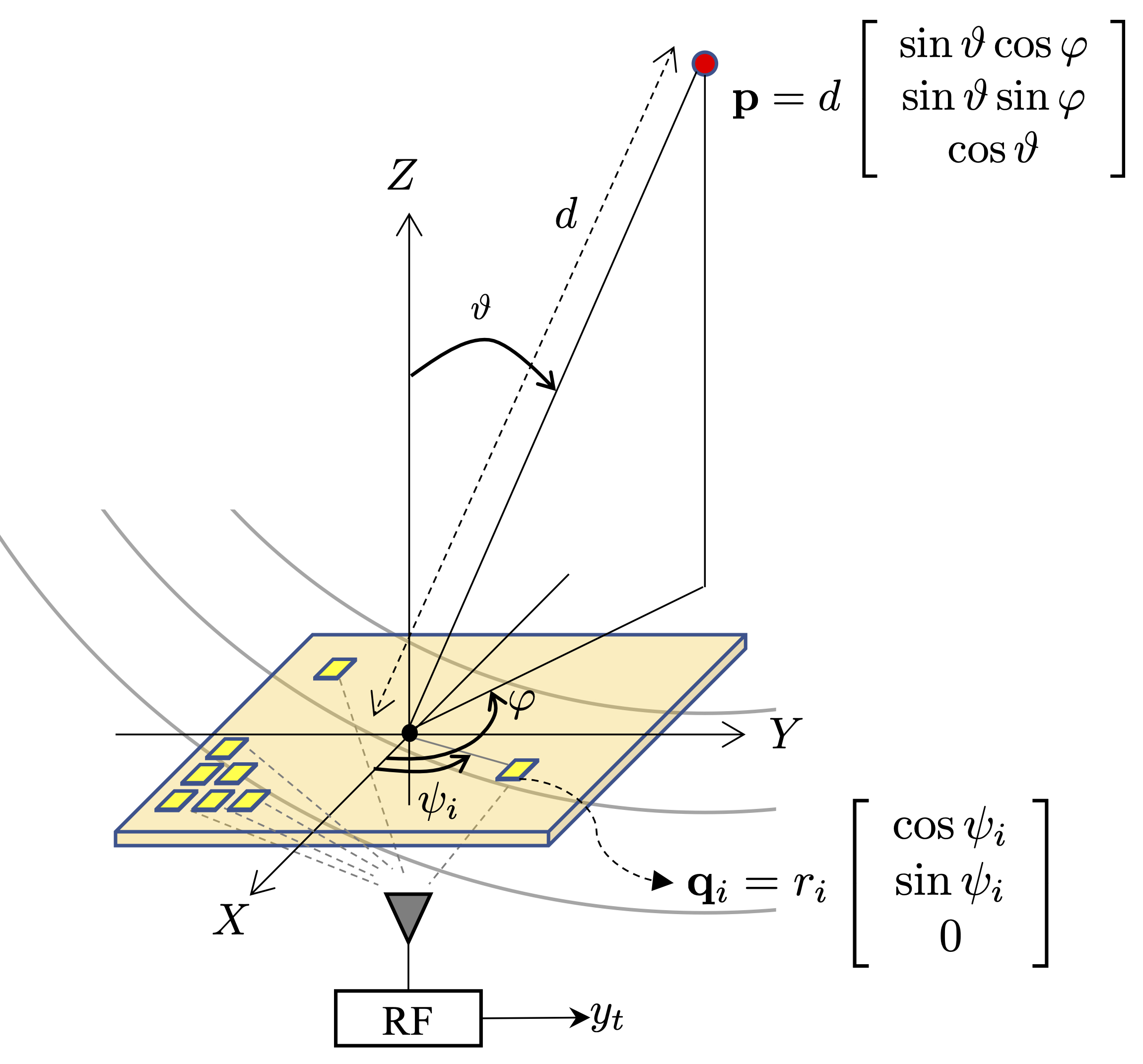}
    \caption{The considered system setup with a single transmitting user and a single receiver comprising of RIS lens and a receive antenna element attached to a reception RF chain. We aim to localize the user in 3D based on the $T$ scalar baseband observations $y_t$,  $\forall$$t=1,2,\ldots,T$. }
    \label{fig:model}
\end{figure}
In this section, we describe the considered system setup for 3D localization with RIS lens and present the channel models that will be used for PEB analysis and performance evaluation. 

\subsection{Geometry Model}
We consider the wireless system setup of Fig.~\ref{fig:model} including a single transmitting user at the location $\bm{p}=[x~y~z]^{\ttt}$ with $x,y\in\mathbb{R}$ and real-valued $z>0$ and an $M$-element RIS lens laying in the XY plane with reference location $\bm{0}_3$, which is placed in close proximity to a single antenna with a respective receive RF chain, located at $\bm{p}_{\mathrm{ant}}\in\mathbb{R}^{3}$. Each $i$-th element of the RIS lens, with $i=1,2,\ldots,M$, is assumed to have the size $A=a\times a$ (typically, less or equal to $\lambda^2/4$ with $\lambda$ being the signal wavelength) and being located at $\bm{q}_{i}=[x_i~y_i~0]^{\ttt}=r_i [\cos \psi_i~\sin \psi_i~0]^\ttt\in\mathbb{R}^{3}$, where $r_i$ denotes the element's distance from the origin and $\psi_i$ is the azimuth angle, as shown in Fig.~\ref{fig:model}. Let $\vartheta \in[0,\pi/2]$ be the angle between the Z axis (i.e., the normal of the RIS) and $\bm{p}$ and $\varphi \in [0,2\pi]$ represent the angle between the projection of 
$\bm{p}$ on the XY plane and the X axis, measured counter-clockwise. We, henceforth, call $\vartheta,\varphi$ the angles-of-arrival (AoAs). Using the latter notations, we  introduce the wavevector as a function of $\vartheta$ and $\varphi$:
\begin{align}
\bm{k}(\vartheta,\varphi) = -\frac{2 \pi}{\lambda}\left[\begin{array}{c}
\sin\vartheta\cos\varphi\\
\sin\vartheta\sin\varphi\\
\cos\vartheta
\end{array}\right],   
\end{align}
which can be used in expressing the unknown position vector as $\bm{p}=-\lambda d\bm{k}(\vartheta,\varphi)/2 \pi$, 
%
%in which $\lambda$ is the signal wavelength, $\vartheta \in[0,\pi/2]$ is the angle between the Z axis (i.e., the normal of the RIS) and $\bm{p}$, and $\varphi \in [0,2\pi]$ is the angle between the projection of 
%$\bm{p}$ on the XY plane and the X axis, measured counter-clockwise. We call $\vartheta,\varphi$ the angles-of-arrival (AoAs). 
%Hence,
%\begin{align}
%\bm{p}=-\frac{\lambda d }{2 \pi}\bm{k}(\bm{p}),
%\end{align}
where $d$ is the Euclidean distance between $\bm{p}$ and the RIS reference location, i.e., $d\triangleq\Vert \bm{p}\Vert$. 
%Since the wavevector is independent of the distance $d$, we can also write $\bm{k}(\vartheta,\varphi)$. 
We model \textit{a priori} information of the user location as a Gaussian pdf $p(\bm{p})$ having the mean $\bm{m}_p\in\mathbb{R}^{3}$ and covariance matrix represented by $\bm{\Sigma}_p\in\mathbb{R}^{3\times3}$. 

\subsection{Signal Model}
We assume a narrowband signal model according to which the transmitter sends the pilot signals $s_{t}$ with $\mathbb{E}\{|s_{t}|^2\}=E_s$ and $t=1,2,\ldots,T$ to the single-antenna receiver via the RIS lens for location estimation. The phase profile of the $M$ RIS lens elements at the time instant $t$ is represented by
$\bm{\Omega}_{t}=\text{diag}(\omega_{t,0},\ldots,\omega_{t,M-1})$, where $|\omega_{t,i}|=1$ $\forall$$i=0,1,\ldots,M-1$. The baseband received signal at the output of the receive RF chain at each time $t$ can be mathematically expressed as 
\begin{align}
y_{t} & = e^{\jmath \theta} \bm{h}_{\mathrm{ant}}^{\ttt}\bm{\Omega}_{t}(\bm{\rho}(\bm{p})\circ\bm{a}(\bm{p}))s_{t}+n_{t},
\end{align}
where 
$\theta  = -\frac{2\pi d}{\lambda} + \theta_{\mathrm{sync}}$,  
%\begin{align}
%    \theta & = -\frac{2\pi d}{\lambda} + \theta_{\mathrm{sync}}\\
%    \rho^2 & =  f(\vartheta,\varphi)\frac{A \cos \vartheta}{4\pi d^2 }, 
%\end{align}
in which $\theta_{\mathrm{sync}}$ is a global phase offset, which accounts for the lack of phase synchronization between transmitter and receiver and all other practical effects such as the phase response of the receive and transmit antennas,
$\bm{h}_{\mathrm{ant}}\in\mathbb{C}^{M\times1}$ includes the fixed and known channel gains from the
RIS lens to the single receive antenna, $\bm{\rho}(\bm{p}) \ge \mathbf{0}_M$ is the vector with the amplitudes of the wireless channels among the RIS elements and the transmitter, and $\bm{a}(\bm{p})\in\mathbb{C}^{M\times1}$ is the vector of channel phases. Notation 
$n_{t}$ is the zero-mean additive Gaussian noise with variance $N_{0}/2$ per real dimension; we assume that $n_{t}$'s are independent and identically distributed.
By introducing $\bm{w}_t=\bm{\Omega}_{t}\bm{h}_{\mathrm{ant}}$, $\bm{W}=[\bm{w}_1, \bm{w}_2, \ldots, \bm{w}_T] \in \mathbb{C}^{M\times T}$, $\bm{s}=[s_1,s_2,\ldots,s_T]^\ttt$, and $\bm{n}=[n_1,n_2,\ldots,n_T]^\ttt$, the measurement vector $\bm{y}=[y_1,y_2,\ldots,y_T]^\ttt$ can be compactly expressed as
\begin{align}
    \bm{y}=e^{\jmath \theta}\text{diag}(\bm{s})\bm{W}^\ttt(\bm{\rho}(\bm{p})\circ\bm{a}(\bm{p})) + \bm{n}. \label{eq:ObsModel}
\end{align}
For simplicity, we will next assume that $s_t=\sqrt{E_s}$, $\forall$$t$. 

\subsection{Channel Model}
Three channel models (CMs) for $\bm{h}_{\mathrm{ant}}$, $\bm{\rho}(\bm{p})$, and $\bm{a}(\bm{p})$ are henceforth considered using a common terminology in communication theory: the channel under the plane-wave model is termed as far-field, while under the curved wave model is termed near-field. We note that from an electromagnetic point of view, both are far-field models\footnote{As mentioned in \cite{friedlander2019localization}, ``the model used in the signal processing literature for near-field localization is in fact the far-field model of electromagnetics or an approximation thereof.''}. Electromagnetic near-field effects within the Fraunhofer distance are ignored in this work. CM1 is the standard far-field model, where the amplitude of the received signal is constant across the RIS elements and the phase depends on AoAs. CM2 is the standard near-field model, where the amplitude is constant and the phase depends on the distance to each RIS element. Finally, CM3 is the improved near-field model from \cite{bjornson2020power}, where the amplitude at each RIS element depends on its location with respect to the user location, and the  phase is as in CM2. The latter models for the involved channels are mathematically described as follows:
\begin{itemize}
    \item \emph{CM1 (standard far-field):} Under this model, it holds:
\begin{align}
    \bm{\rho}(\bm{p}) & = \rho \bm{1}_M\\ \label{eq_cm1_ap}
    [\bm{a}(\bm{p})]_i & =  \exp(-\jmath\bm{q}^{\ttt}_{i}\bm{k}(\vartheta,\varphi)),
\end{align}
where $\rho^2  =  f(\vartheta,\varphi)\frac{A \cos \vartheta}{4\pi d^2 }$ represents the common power for all RIS elements and  $f(\vartheta,\varphi)$ is a correction factor (see later, in CM3).%\GA{shall we give the definition here?}). 
\item \emph{CM2 (standard near-field):} Under this model, $\bm{\rho}(\bm{p}) = \rho \bm{1}_M$ as in CM1, but the phase accounts for wavefront curvature and is given by
\begin{align}
    [\bm{a}(\bm{p})]_i & =  \exp\left(-\jmath\frac{2\pi}{\lambda}\left(\Vert\bm{p}-\bm{q}_{i}\Vert-d\right)\right). \label{eq:near-field-phase}
\end{align}
It is readily verified that when $d \gg \Vert \bm{q}_i \Vert$, $\forall i$, then CM2 reverts to CM1.\footnote{To see this, note that  $\Vert\bm{p}-\bm{q}_{i}\Vert-\Vert\bm{p}\Vert=d(1+{r_{i}^{2}}/{d^{2}}-2{r_{i}}/{d}\sin\vartheta\cos(\varphi-\psi_{i}))^{1/2}-d$, where $r_{i}$ and $\psi_{i}$ are defined in Fig.~\ref{fig:model}. For $d\gg r_i$, this becomes $-r_i\sin\vartheta\cos(\varphi-\psi_{i})$, which does not depend on $d$. } 
\item \emph{CM3 (improved near-field):} The amplitudes $\bm{\rho}(\bm{p})$ can be upper bounded  according to \cite[Lemma 1]{bjornson2020power} as
\begin{align}
    & [\bm{\rho}(\bm{p})]^2_i = \left(12\pi\right)^{-1} \label{eq:channelModelNearField}\\ & \times \sum_{\substack{\mathsf{x}\in \mathcal{X}_i\\\mathsf{y}\in \mathcal{Y}_i}}\frac{\mathsf{x}\mathsf{y}}{ (\mathsf{y}^2+z^2)g(\mathsf{x},\mathsf{y})}+2\arctan\left(\frac{\mathsf{x}\mathsf{y}}{z^2g(\mathsf{x},\mathsf{y})}\right),\notag
\end{align}
in which $\mathcal{X}_i=\{a/2+(x_i-x), a/2-(x_i-x)\}$, $\mathcal{Y}_i=\{a/2+(y_i-y), a/2-(y_i-y)\}$, and $g(\mathsf{x},\mathsf{y})=\sqrt{\mathsf{x}^2/z^2 + \mathsf{y}^2/z^2 +1}$. In order to ensure consistency between the far-field and near-field models, we have found that $f(\vartheta,\varphi)=1-\sin^2(\vartheta)\sin^2(\varphi)$, which accounts for the specific linear polarization and the associated polarization loss considered in \cite{bjornson2020power}.\footnote{More specifically, the model in \cite{bjornson2020power} assumes that the transmitter excites only the component of the electric field along the Y-axis (this is one of the two axis where the RIS lies). Note that this  assumption  cannot be valid for arbitrary user orientations, so the model should be generalized. This is beyond the scope of the current paper.} Then, for $d \gg \Vert \bm{q}_i \Vert$, $\forall i$, CM3 reverts to CM2, which in turn reverts to CM1. 
\end{itemize}

%\footnote{More specifically, the model in \cite{bjornson2020power} assumes that the transmitter excites only the component of the electric field and this happens only along the Y-axis (this is one of the two axis where the RIS lies). Note that this  assumption  cannot be valid for all user locations, so that the model should be generalized. This is beyond the scope of the current paper.}

These latter models are used throughout this paper, as follows. CM3 is used to generate $\bm{h}_{\mathrm{ant}}$, and also the actual unknown channel in the performance evaluation. In the Fisher information analysis, phase profile design, and algorithm derivation, CM1 and CM2 will be considered. %In the performance evaluation, the unknown channel will be generated according to CM3. 
%\FK{FIM analysis is valid for both CM1 and CM2 as the phase function is generic. For algorithm derivation, we use CM1 phase profile in (5). For phase profile design, both CM1 and CM2 are used.}

\section{Fisher Information Analysis}

\subsection{Introduction}
The observations in baseband for the $T$ transmitted pilots at the receiver under CM2 can be expressed using \eqref{eq:ObsModel} as 
\begin{align}\label{eq:observations}
     \bm{y}=\sqrt{E_s}\rho e^{\jmath \theta} \bm{W}^\ttt\bm{a}(\bm{p}) + \bm{n}.
\end{align}
Introducing the noise-free observation $\bm{\mu}=\sqrt{E_s}\rho e^{\jmath \theta} \bm{W}^\ttt\bm{a}(\bm{p})$ and the $5\times1$ vector of unknowns $\bm{\eta}=[\rho~\theta~\bm{p}^\ttt]^\ttt$, the Fisher information matrix (FIM) $\bm{J}(\bm{\eta}) \in \mathbb{R}^{5 \times 5}$ is defined as \cite{VanTrees}
\begin{align}\label{eq:PEB_definition}
    \bm{J}(\bm{\eta})=\frac{2}{N_0}\Re \left\{\left(\frac{\partial \bm{\mu}}{\partial\bm{\eta}}\right)^{\text{H}} \frac{\partial\bm{\mu}}{\partial\bm{\eta}}\right\}.
\end{align}
Using the latter expression, the PEB in meters is given by 
\begin{align}
    \mathrm{PEB}=\sqrt{\mathrm{trace}([\bm{J}^{-1}(\bm{\eta})]_{3:5,3:5})},
\end{align}
and can be related to the root mean squared error (RMSE) of any unbiased estimator $\hat{\bm{p}}$ by the inequality:
\begin{align}
    \mathrm{PEB} \le \mathrm{RMSE}\triangleq \sqrt{\mathbb{E}\{\Vert \hat{\bm{p}}-\bm{p}\Vert^2 \}}.
\end{align}

\subsection{PEB Derivation}
We can write the partial derivatives needed in \eqref{eq:PEB_definition} as ${\partial\bm{\mu}}/{\partial\rho} =\sqrt{E_s} e^{\jmath \theta} \bm{W}^\ttt\bm{a}(\bm{p})$, ${\partial\bm{\mu}}/{\partial\theta}=\sqrt{E_s}\rho \jmath e^{\jmath \theta} \bm{W}^\ttt\bm{a}(\bm{p})$, and ${\partial\bm{\mu}}/{\partial\bm{p}} 
 =\sqrt{E_s}\rho e^{\jmath \theta} \bm{W}^\ttt\bm{D}(\bm{p})$,  
%\begin{align}
 %   \frac{\partial\bm{\mu}}{\partial\rho}& =\sqrt{E_s} e^{\jmath \theta} \bm{W}^\ttt\bm{a}(\bm{p})\\
  %  \frac{\partial\bm{\mu}}{\partial\theta}& =\sqrt{E_s}\rho \jmath e^{\jmath \theta} \bm{W}^\ttt\bm{a}(\bm{p})\\
   % \frac{\partial\bm{\mu}}{\partial\bm{p}} & =\sqrt{E_s}\rho e^{\jmath \theta} \bm{W}^\ttt\bm{D}(\bm{p}) 
%\end{align}
where $\bm{D}(\bm{p})=\frac{\partial\bm{a}(\bm{p})}{\partial\bm{p}}\in\mathbb{C}^{M\times3}$ which is computed as
\begin{align}
\frac{\partial\bm{a}(\bm{p})}{\partial\bm{p}}=\jmath\frac{2\pi}{\lambda}\left(\text{diag}(\bm{a}(\bm{p}))\bm{K}^\ttt+\bm{a}(\bm{p})\frac{\bm{p}^\ttt}{d}\right).
\end{align}
In this expression, $\bm{K}=[\bm{e}_{0},\bm{e}_{1},\ldots,\bm{e}_{M-1}]$
with $\bm{e}_{i}=(\bm{q}_{i}-\bm{p})/\Vert\bm{q}_{i}-\bm{p}\Vert$.
Introducing the positive semidefinite matrix $\bm{F} = \WW^{*} \WW^\ttt$, the non-zero entries of the  FIM are given as follows. For its diagonal elements holds:
\begin{align}
[\bm{J}(\bm{\eta})]_{1,1}& =\frac{2E_{s}}{N_{0}} \bm{a}^{\text{H}}(\bm{p})\bm{F}\bm{a}(\bm{p}),\\
[\bm{J}(\bm{\eta})]_{2,2}& =\frac{2E_{s}\rho^2}{N_{0}} \bm{a}^{\text{H}}(\bm{p})\bm{F}\bm{a}(\bm{p}), \\
[\bm{J}(\bm{\eta})]_{3:5,3:5}& =\frac{2E_{s}\rho^2}{N_{0}} \Re\left(\bm{D}^{\text{H}}(\bm{p})\bm{F}\bm{D}(\bm{p})\right),
\end{align}
and for the off-diagonal elements (above the main diagonal):
\begin{align}
    [\bm{J}(\bm{\eta})]_{1:2,3:5} & =\frac{2E_{s}\rho}{N_{0}}  \left[\begin{array}{c} \Re\left(\bm{a}^{\text{H}}(\bm{p})\bm{F}\bm{D}(\bm{p})\right) \\ \rho\Im\left(\bm{a}^{\text{H}}(\bm{p})\bm{F}\bm{D}(\bm{p})\right) \end{array}\right].
\end{align}
%\begin{align}
%& = \frac{2E_{s}}{N_{0}}\left[\begin{array}{ccc}
%\bm{a}^{\text{H}}(\bm{p})\bm{F}\bm{a}(\bm{p}) & 0 & \rho\Re\left(\bm{a}^{\text{H}}(\bm{p})\bm{F}\bm{D}(\bm{p})\right)\\
%0 & \rho^{2}\bm{a}^{\text{H}}(\bm{p})\bm{F}\bm{a}(\bm{p}) & \rho^{2}\Im\left(\bm{a}^{\text{H}}(\bm{p})\bm{F}\bm{D}(\bm{p})\right)\\
%\rho\Re\left(\bm{D}^{\text{H}}(\bm{p})\bm{F}\bm{a}(\bm{p})\right) & -\rho^{2}\Im\left(\bm{D}^{\text{H}}(\bm{p})\bm{F}\bm{a}(\bm{p})\right) & \rho^{2}\Re\left(\bm{D}^{\text{H}}(\bm{p})\bm{F}\bm{D}(\bm{p})\right)
%\end{array}\right]
%\end{align}

To obtain further insights on the PEB performance, we derive the equivalent FIM of the user location  as \cite{shen2010fundamental}
\begin{align} \label{eq_fim_loc}
& \bm{J}(\bm{p})  = [\bm{J}(\bm{\eta})]_{3:5,3:5}\\
&-[\bm{J}(\bm{\eta})]_{3:5,1:2} [\bm{J}(\bm{\eta})]^{-1}_{1:2,1:2} [\bm{J}(\bm{\eta})]_{1:2,3:5} \notag\\
    & =\frac{2\rho^{2}E_{s}}{N_{0}}\Re\left\{ \bm{D}^{\text{H}}(\bm{p})\left[\bm{F}-\frac{\bm{F}\bm{a}(\bm{p})\bm{a}^{\text{H}}(\bm{p})\bm{F}}{\bm{a}^{\text{H}}(\bm{p})\bm{F}\bm{a}(\bm{p})}\right]\bm{D}(\bm{p})\right\}. \notag
\end{align} 
By defining $\EE = \WW^\ttt \DD$ and $\bb = \WW^\ttt \aaa$, we obtain
\begin{align}\label{eq:FIMprojection}
    \bm{J}(\bm{p}) &= \frac{2\rho^{2}E_{s}}{N_{0}}\Re\left\{\EE^\hht \left(\bm{I}-\frac{\bb \bb^\hht}{ \Vert \bb \Vert^2}\right) \EE \right\}\\ \nonumber
    = \frac{2\rho^{2}E_{s}}{N_{0}}&\Re\left\{ \EE^\hht \projnull{\bb} \EE \right\} = \frac{2\rho^{2}E_{s}}{N_{0}}\Re\left\{ (\projnull{\bb} \EE  )^\hht \projnull{\bb} \EE \right\},  
\end{align}
where $\projnull{\bb} \EE$ represents the component of $\WW^\ttt \DD$ that lies in the subspace orthogonal to $\bb =\WW^\ttt \aaa$; $\projnull{\bb}$ is the orthogonal projection operator over $\bb$. Thus, the PEB can be expressed in a compact form up to an SNR scaling factor as $\mathrm{PEB} = \sqrt{\mathrm{trace}(  \Re\left\{ \EE^\hht \projnull{\bb} \EE \right\} ^{-1})}$. 
%\begin{align}
%    \mathrm{PEB} = \sqrt{\mathrm{trace}\left(  \Re\left\{ \EE^\hht \projnull{\bb} \EE \right\} ^{-1}\right)}% \Vert \projnull{\bb} \EE^\hht \Vert_F.
%\end{align}
%
Obviously, if the derivative $\DD$ is almost orthogonal to $\aaa$ (after projection through $\WW^\ttt$), then a large amount of positional information is available resulting in small PEB. In other words, the position estimation accuracy depends on how well the RIS phase profile can distinguish the steering vector and its derivative. In addition, when $d \gg r_i$ (i.e., under the plane wave model), it holds $\text{diag}(\bm{a}(\bm{p}))\bm{K}^\ttt \to -\bm{a}(\bm{p}){\bm{p}^\ttt}/{d}$, and thus $\bm{J}(\bm{p}) \to 0$. This indicates that the PEB increases further away from the RIS-based lens, irrespective of the path loss.

%\FK{An idea for simplification: Note that $\bm{F} = \WW^{*} \WW^\ttt$. Then, we can write the argument in \eqref{eq_fim_loc} as
%\begin{align}
 %   \bm{J}_{\rm{arg}} = \DDh \WW^{*} \WW^\ttt \DD - \frac{ \DDh \WW^{*} \WW^\ttt \aaa \aaah \WW^{*} \WW^\ttt \DD }{\aaah \WW^{*} \WW^\ttt \aaa} ~.
%\end{align} 
%Define $\EE \triangleq \DDh \WW^{*}$ and $\bb = \WW^\ttt \aaa$. We obtain
%\begin{align}
 %   \bm{J}_{\rm{arg}} &= \EE \EE^\hht - \frac{ \EE \bb \bb^\hht \EE^\hht }{ \Vert \bb \Vert^2 } = \EE \projnull{\bb} \EE^\hht \\
  %  &= \big(\projnull{\bb} \EE^\hht \big)^\hht \big(\projnull{\bb} \EE^\hht\big) ~.
%\end{align}
%Notice that $\projnull{\bb} \EE^\hht$ represents the component of $\WW^\ttt \DD$ that lies in the subspace orthogonal to $\WW^\ttt \aaa$. Consider the trace of the EFIM:
%\begin{align}
 %   \rm{tr}(\bm{J}_{\rm{arg}}) = \Vert \projnull{\bb} \EE^\hht \Vert_F^2
%\end{align}
%Intuitively, if the derivative $\DD$ is almost orthogonal to $\aaa$ (after projection through the RIS codebook $\WW^\ttt$), then we have large amount of positional information. Hence, the estimation accuracy depends on how well the codebook can distinguish the steering vector and its derivative.}

%In case a priori information is available, the FIM becomes
%\begin{align}
%\bm{J}^{\mathrm{post}}(\bm{p}) =   %\bm{J}(\bm{p})+\bm{\Sigma}^{-1},
%\end{align}
%where $\bm{\Sigma}$ is the a priori covariance. 

%Finally, to PEB (expressed in meters) is 
%\begin{align}
%    \mathrm{PEB}=\sqrt{\mathrm{trace}(\bm{J}^{-1}(\bm{p}))}.
%\end{align}

\section{RIS Phase Profile Design} \label{sec:RISConfigurations}
The performance of the location estimation depends on the choice of RIS phase profiles $\bm{\Omega}_t$. In order to remove the effect of the phases in $\bm{h}_{\mathrm{ant}}$, we set $\bm{\Omega}_t=\bm{\Omega}_{\mathrm{ant}}\tilde{\bm{\Omega}}_t$, where the fixed phases  $\bm{\Omega}_{\mathrm{ant}}=\text{diag}(\omega_{\mathrm{ant},0},\ldots,\omega_{\mathrm{ant},M-1})$  ensure that $[\bm{h}^\ttt_{\mathrm{ant}}\bm{\Omega}_{\mathrm{ant}}]_i = |[\bm{h}^\ttt_{\mathrm{ant}}]_i|,\,\forall i$ \cite{guidi2019radio}, exploiting the knowledge of $\bm{h}_{\mathrm{ant}}$. It then remains to design $\tilde{\bm{\Omega}}_t=\text{diag}(\tilde{\omega}_{t,0},\ldots,\tilde{\omega}_{t,M-1})$. We consider the following three designs for $\tilde{\omega}_{t,i}$ $\forall$$i,t$:
\begin{itemize}
    \item \emph{Random:} In this approach, we set  $\tilde{\omega}_{t,i}=\exp(\jmath \psi_{t,i})$, where $\psi_{t,i} \sim \mathcal{U}(0,2\pi)$ independently for each $i$-th RIS element and each time instant $t$. 
    \item \emph{Directional:} We set each phase configuration under CM1 as
    $\tilde{\omega}_{t,i} =\exp(+\jmath\bm{q}^{\ttt}_{i}\bm{k}(\vartheta^{(k)},\varphi^{(k)}))$, where the samples $\vartheta^{(k)}$ and $\varphi^{(k)}$ are obtained from the \textit{a priori} pdf $p(\bm{p})$. 
    \item \emph{Positional:}
    Under CM2, we set
    $\tilde{\omega}_{t,i} =\exp\left(+\jmath\frac{2\pi}{\lambda}\left(\Vert\bm{p}^{(k)}-\bm{q}_{i}\Vert-d^{(k)}\right)\right)$, where $\bm{p}^{(k)}$ and $d^{(k)}$  are sampled from $p(\bm{p})$. 
\end{itemize}

\begin{figure*}
    \centering
    \includegraphics[width=0.9\textwidth]{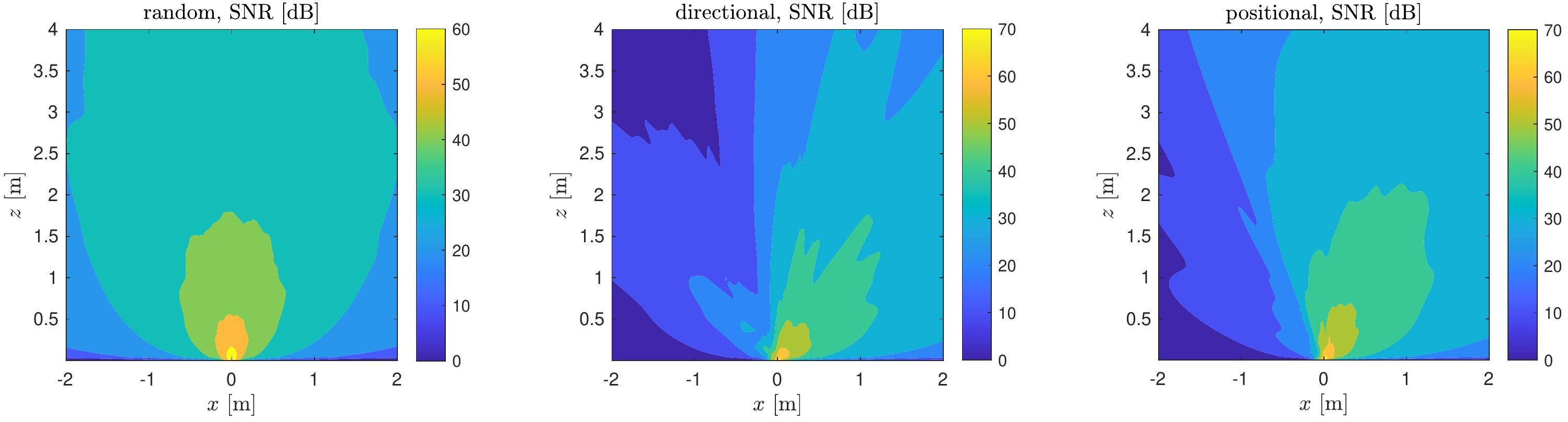}
    \caption{SNR from \eqref{eq:SNR_def} in dB in the $X=Y$ plane for random, directional, and positional RIS phase profiles, with $\bm{m}_p=[0.1\,0.1\,0.1]^\ttt$ and $\bm{\Sigma}_p=0.01 \bm{I}_3$. For visualization purposes, the SNR values are truncated at 0 dB. Infinite resolution of the phase profiles is assumed, though with 2 or 3 bits, the SNR is close to the results in the figure. }
    \label{fig:SNRfromRIS}
\end{figure*}

To understand the difference between the different phase profiles, we show the SNR as a function of the location in the plane $Y=X$ for a single realization of a phase profile for each of the three choices. The SNR is defined as
\begin{align}
     \text{SNR}=\frac{1}{T}\sum_{t=1}^{T}\frac{{E_s}\rho^2}{N_0} |\bm{w}_t^\ttt\bm{a}(\bm{p})|^2. \label{eq:SNR_def}
\end{align}
The results are shown in Fig.~\ref{fig:SNRfromRIS}, assuming an \textit{a priori} position distribution with mean $[0.1\,0.1\,0.1]^\ttt$ and covariance $0.01 \bm{I}_3$. We observe that the random case leads to uniform SNR for all locations subject to path loss, with reduced values close to the end-fire of the RIS. For the directional case, a higher SNR is achieved along the chosen direction, with reduced SNR in other locations, as compared to the random case. 
Finally, for the positional case, beams tend to be slightly broader.

\section{Location Estimation}
\subsection{Maximum Likelihood Estimator}
We introduce $\alpha=\rho e^{j\theta}$ so that the maximum likelihood estimate of the channel gain and user location is given by 
\begin{align}
[\hat{\alpha},\hat{\bm{p}}] & = \arg\max_{\alpha,\bm{p}}\,f(\bm{y}|\alpha,\bm{p})\\ & =\arg\min_{\alpha,\bm{p}}\,\Vert \bm{y}-\sqrt{E_s}\alpha \bm{W}^\ttt\bm{a}(\bm{p})\Vert^2. \label{eq:MLProblem}
\end{align}
Solving for $\alpha$, yields the gain estimation as a function of $\bm{p}$:
\begin{align}\label{eq:alpha_est}
    \hat{\alpha}(\bm{p})=\frac{\bm{a}^\mathrm{H}(\bm{p})\bm{W}^*\bm{y}}{\sqrt{E_s}\Vert  \bm{W}^\ttt\bm{a}(\bm{p}) \Vert ^2}.
\end{align}
Consequently, the estimate $\bm{p}$ can be calculated as: 
\begin{align}
    \hat{\bm{p}}& =\arg \min_{\bm{p}} \Vert \bm{y} -  \sqrt{E_s}\hat{\alpha}(\bm{p}) \bm{W}^\ttt\bm{a}(\bm{p})\Vert^2. \label{eq:positionOpt}\\
     & = \arg \min_{\bm{p}} \Vert \projnull{\bb}\bm{y}\Vert^2,
\end{align}
where $\projnull{\bb}$ is the orthogonal projection operator over $\bb =\WW^\ttt \aaa$, as defined in \eqref{eq:FIMprojection}.
%We introduce $\bm{W}=[\bm{w}_1, \bm{w}_2, \ldots, \bm{w}_T] \in \mathbb{C}^{M\times T}$ and 
%$\bm{\gamma}=E_s\bm{W}^\ttt \bm{a}(\bm{p}) \in \mathbb{C}^{T \times 1}$. 
%Solving for $\text{\ensuremath{\alpha}}$ yields 
%\begin{align}
 %   \hat{\alpha}(\bm{p})=\frac{\bm{\gamma}^\mathrm{H}\bm{y}}{\Vert \bm{\gamma} \Vert ^2}
%\end{align}
%so that we can find 
%\begin{align}
 %   \hat{\bm{p}}& =\arg \max_{\bm{p}} \Lambda(\bm{p}) \label{eq:positionOpt}\\
  %  \Lambda(\bm{p}) & = - \Vert \bm{y} -\hat{\alpha}(\bm{p})  \bm{\gamma}\Vert^2 
%\end{align}

\subsection{Low Complexity Localization}
To solve \eqref{eq:positionOpt}, we make use of the underlying structure of the optimization problem in spherical coordinates, which leads to a three-stage estimator, as follows. We first express each $i$-th exponential term in \eqref{eq_cm1_ap} as
\begin{align}\label{eq_cm1_ap2}
    [\bm{a}(\vartheta, \varphi)]_i = \exp\left(-\jmath \frac{2 \pi}{\lambda} r_i \sin(\vartheta) \cos(\varphi - \psi_i) \right). 
\end{align}
Similar to \cite{wang2019super}, we employ the Jacobi-Anger expansion to re-express \eqref{eq_cm1_ap2} as
\begin{align}
    [\bm{a}(\vartheta, \varphi)]_i = \sum_{n=-\infty}^{\infty} \jmath^n J_n\left(-\frac{2 \pi}{\lambda} r_i \sin(\vartheta)\right) e^{\jmath n (\varphi - \psi_i)},
\end{align}
where $J_n(\cdot)$ is the $n$-th order Bessel function of the first kind. Neglecting the terms with $\lvert n \rvert > N$ for a given $N$ (note that $\lvert J_n(\cdot) \rvert$ decays to zero as $\lvert n \rvert$ increases), we obtain
\begin{align}
    [\bm{a}(\vartheta, \varphi)]_i \approx \sum_{n=-N}^{N} \jmath^n J_n\left(-\frac{2 \pi}{\lambda} r_i \sin(\vartheta)\right) e^{\jmath n (\varphi - \psi_i)} ~. \label{eq:expansion}
\end{align}
By defining the following parameters for $n = -N, \ldots, N$:
\begin{align}
    \left[ \bm{g}_i(\vartheta) \right]_n &= \jmath^n J_n\left(-\frac{2 \pi}{\lambda} r_i \sin(\vartheta)\right) e^{-\jmath n  \psi_i} \\ \label{eq_h_varphi}
    \left[ \bm{h}(\varphi) \right]_n &= e^{\jmath n \varphi },
\end{align}
yields $ [\bm{a}(\vartheta, \varphi)]_i = \bm{g}_i^\ttt(\vartheta) \bm{h}(\varphi)$. 
%\begin{align}
 %   [\bm{a}(\vartheta, \varphi)]_i = \bm{g}_i^\ttt(\vartheta) \bm{h}(\varphi) ~. 
%\end{align}
It can be easily verified that $\bm{a}(\vartheta, \varphi) \approx \bm{G}^\ttt(\vartheta) \bm{h}(\varphi)$, where $\bm{G}(\vartheta) = \left[ \bm{g}_0(\vartheta) \, \ldots \, \bm{g}_{M-1}(\vartheta)  \right]$.
%Defining $\bm{G}(\vartheta) = \left[ \bm{g}_0(\vartheta) \, \ldots \, \bm{g}_{M-1}(\vartheta)  \right]$, then we have $\bm{a}(\vartheta, \varphi) \approx \bm{G}^\ttt(\vartheta) \bm{h}(\varphi)$. 
%\begin{align}
%    \bm{a}(\vartheta, \varphi) = \bm{G}^\ttt(\vartheta) \bm{h}(\varphi) ~.
%\end{align}
Now, the angular steering vector $\bm{a}(\vartheta, \varphi)$ has a form that is separable in the angles $\vartheta$ and $\varphi$. 

We are now ready to proceed with our three-stage estimator. 
\begin{enumerate}
    \item \emph{Estimation of $\vartheta$:} Under CM1, we rewrite $\bm{y}$ in \eqref{eq:observations} as
    \begin{align}\label{eq:y_theta}
        \bm{y}=\sqrt{E_s}\alpha \bm{W}^\ttt\bm{G}^\ttt(\vartheta) \bm{h}(\varphi) + \bm{n}.
    \end{align}
    By introducing the unstructured vector $\bm{v} =\sqrt{E}_s\alpha \bm{h}(\varphi)$, the estimate of $\bm{v}$ can be expressed by the following function of $\vartheta$: $\hat{\bm{v}}(\vartheta)=\left( \bm{G}^*(\vartheta)\bm{W}^*\right)^\dagger    \bm{G}^*(\vartheta)\bm{W}^* \bm{y}$.
    %\begin{align}
     %   \hat{\bm{v}}(\vartheta)=\left( \bm{G}^*(\vartheta)\bm{W}^*\right)^\dagger    \bm{G}^*(\vartheta)\bm{W}^* \bm{y}. \end{align}
    Hence, the estimation for angle $\vartheta$ can be obtained as
        \begin{align}
            \hat{\vartheta}=\arg \min_{\vartheta} \Vert \bm{y} -\bm{W}^\ttt\bm{G}^\ttt(\vartheta) \hat{\bm{v}}(\vartheta)\Vert^2,
        \end{align}
        which can be solved with a simple line search. 
        \item \emph{Estimation of $\varphi$:} Again, under CM1 and using the estimate $\hat{\vartheta}$, expression \eqref{eq:y_theta} can be written as $\bm{y}=\sqrt{E_s}\alpha \bm{W}^\ttt\bm{G}^\ttt(\hat{\vartheta}) \bm{h}(\varphi) + \bm{n}$. We estimate $\alpha$ for each value of $\varphi$ similarly as in (\ref{eq:alpha_est}), but replacing $\bm{a}(\bm{p})$ with $\bm{G}^\ttt(\hat{\vartheta}) \bm{h}(\varphi)$, leading to $\hat{\alpha}(\varphi)$. Then, we can solve for angle $\varphi$ as follows:
        \begin{align}
            \hat{\varphi}=\arg \min_{\varphi} \Vert \bm{y} -\sqrt{E}_s\hat{\alpha}(\varphi)\bm{W}^\ttt\bm{G}^\ttt(\hat{\vartheta}) {\bm{h}}(\varphi)\Vert^2,
        \end{align}
        which requires a second line search.
        \item  \emph{Estimation of $d$:} Under CM2, given the estimates $\hat{\vartheta}$ and $\hat{\varphi}$, we introduce $\bm{p}(d)=d [\sin\hat{\vartheta}\cos\hat{\varphi}\,\,\sin\hat{\vartheta}\sin\hat{\varphi}\,\,\cos\hat{\vartheta}]^\top$, 
    %\begin{align}
%\bm{p}(d) = d\left[\begin{array}{c}
%\sin\hat{\vartheta}\cos\hat{\varphi}\\
%\sin\hat{\vartheta}\sin\hat{\varphi}\\
%\cos\hat{\vartheta}
%\end{array}\right],   
%\end{align}
 from which we determine $\hat{\alpha}(\bm{p}(d))$, an unstructured estimate of $\alpha$ as in (\ref{eq:alpha_est}), and finally solve the optimization:
\begin{align}
    \hat{d}& =\arg \min_d  
    \Vert \bm{y} -  \sqrt{E_s}\hat{\alpha}(\bm{p}(d)) \bm{W}^\ttt\bm{a}(\bm{p}(d))\Vert^2,
    %\Vert\bm{y} -\hat{\alpha}(\bm{p}(d))  \bm{\gamma}(\bm{p}(d))\Vert^2.
\end{align}
which requires a third and final line search. 
\end{enumerate}

\section{Numerical Results}

\subsection{Simulation Setup}
We consider a RIS with $M=2500$ elements (i.e., $50 \times 50$) at $28$ GHz with $\lambda/2$ spacing and area $A=\lambda^2/4$. The single receive antenna is placed behind the RIS lens at $[0~0~-\lambda]^\ttt$. 
The transmit power is $1$ mW, the noise power spectral density is set to $-174~\text{dBm}/\text{Hz}$ and the reception noise figure to 8 dB. We set the number of time instants to $T=200$ and the bandwidth to $1$ MHz, so that localization is based on a $0.2$ ms observation. 
We consider a user with wavevector $\bm{k}$ along the direction $[1~1~1]^\ttt$.  The channels $\bm{h}_{\mathrm{ant}}$ and $\bm{\rho}(\bm{p})\circ\bm{a}(\bm{p})$  in  \eqref{eq:ObsModel} are generated according to CM3, as defined in \eqref{eq:near-field-phase} and \eqref{eq:channelModelNearField}.
\textit{A priori} information of the user location is of the form of a Gaussian pdf $p(\bm{p})=\mathcal{N}(\bm{p};\bm{m}_p,\bm{\Sigma}_p)$ with $\bm{\Sigma}_p=\sigma^2 \bm{I}_3$. This \textit{a priori} information is only used to design the RIS phase profiles, not during localization nor in the PEB calculation. The three RIS profiles designs from Section \ref{sec:RISConfigurations} will be evaluated for $\sigma \in \{0.1,1\}$ m. The channel estimator uses $N=5$ in the expansion \eqref{eq:expansion}.
%unknowns (see vector $\bm{\eta}$ in Section III). and is referred to as $3 \times 1\mathrm{D}$. For comparison purposes, we also consider a two-stage estimator, which directly substitutes \eqref{eq_cm1_ap2} into \eqref{eq:MLProblem} to obtain a joint estimate of $\hat{\vartheta}$ and $\hat{\varphi}$, after which we search for the distance $d$. The latter estimator is referred to as $2\mathrm{D} + 1\mathrm{D}$. 

\subsection{PEB Evaluation}
The PEB as a function of distance in meters is illustrated in Fig.~\ref{fig:PEB} for the three selected RIS phase profile designs and different values of $\sigma$. As a reference, the PEB corresponding to the prior is shown as a horizontal line. We observe that, even with the simple randomized phase profile, relatively low PEB values can be attained, below 1 m for user location distances up to 10 m from the RIS lens. With directional or positional phase profiles, the PEB can be substantially reduced. The positional phase profile performs slightly better than the directional phase profile, but the difference is negligible. Better \textit{a priori} information (i.e., smaller $\sigma$) leads to better PEB. Note that for $\sigma=0.1$ m, the PEB due to the RIS measurements is only better than the \textit{a priori} PEB below 10 m. 

%\begin{rem}[Finite resolution RIS phase profiles]
%While not reported here, we have confirmed that the PEB with randomized phase profiles remains unchanged when using finite resolution RIS codebooks, with 1 bit per phase shifter (i.e., $\omega_{\mathrm{ant},i}\times \tilde{\omega}_{t,i}\in \{-1,+1\}$). With directional and positional profiles, a very small performance degradation is visible. These observations could be important for practical implementations. 
%\end{rem}

\subsection{Localization Accuracy}
\begin{figure}
    \centering
    \input{Figures/RMSEvdistance.tex}
    \caption{RMSE as a function of distance to the RIS lens. }
    \label{fig:RMSE}
\end{figure}
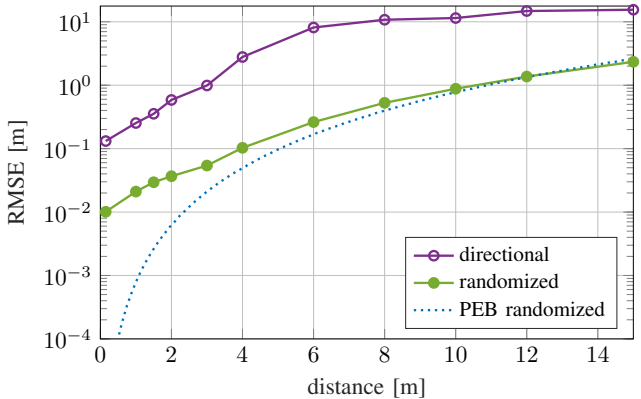

In Fig.~\ref{fig:RMSE}, we show the RMSE of the proposed three-stage localization algorithm,% ($3 \times 1\mathrm{D}$) and the two-stage algorithm ($2\mathrm{D}+1\mathrm{D}$), 
  as a function of distance in meters to the RIS. Since we use  a finite resolution in the angle and delay domain searches (360 bins for $\varphi$, 90 for $\vartheta$ and 500 for $d$), combined with the fact that the CM1 model assumed in the first two stages of the algorithm does not hold for small distances, we are unable to attain the PEB in that regime. With this in mind, the performance of the randomized codebook is close to the PEB, leading to sub-meter localization RMSE within 10 meters from the RIS lens. Paradoxically, the performance under the directional RIS phase profile (shown only for $\sigma = 0.1$ m) is far worse than predicted by the bounds. This can be explained as follows: the directional phase profiles focus energy in the direction of the user, which leads to $\Vert \bm{W}^\ttt\bm{a}(\bm{p})\Vert^2 \approx 0$ for most locations $\bm{p}$ different from the true location. Hence, the objective function \eqref{eq:positionOpt} is nearly flat everywhere, with very narrow peaks around the true position. Due to the finite resolution of the proposed estimators, we miss this peak with high probability, leading to outliers and a degraded RMSE.

\begin{figure}
    \centering
    \input{Figures/PEBvdistance.tex}
    \caption{PEB as a function of distance to the RIS lens. }
    \label{fig:PEB}
\end{figure}
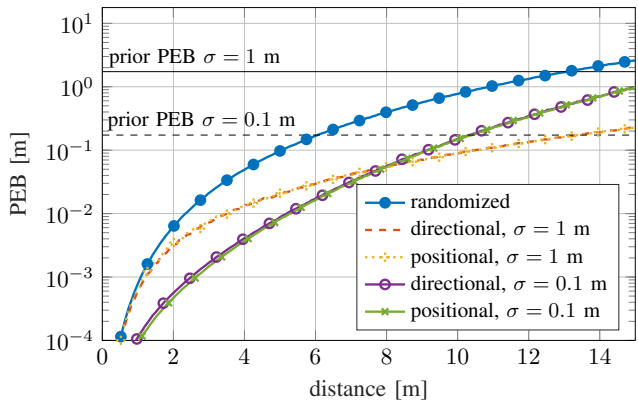

\section{Conclusions}

We considered the problem of localizing a transmitter in 3D using a RIS-based lens and a single receive antenna with a respective RF chain. By exploiting the wavefront curvature, the user location can be estimated, provided that several RIS phase configurations are used. A Fisher information analysis provides insight into the design of these phase configurations. We have also presented a low-complexity 3D localization algorithm, which transforms the 3D problem into 3 one-dimensional problems. Simulation results confirm the validity of the approach and highlight that RIS configurations optimized for localization performance may suffer from degraded performance when not complemented with high-resolution estimators. Based on this, we recommend random RIS phase configurations when low complexity estimation is targeted. 
There are several avenues for further research, including using the amplitude of the received signal for localization, and the inclusion of multi-path and multi-user localization. 

%\HW{To be written.}
%\begin{itemize}
%    \item Can phase profile be designed to simplify estimation? 
 %   \item Should we use amplitude estimation for positioning? Can be done under CM1, CM2, and CM3 (but would be very complicated).
 %   \item What happens under multipath? 
%\end{itemize}

\scriptsize{
\section*{Acknowledgments}
This work was supported, in part, by the Swedish Research Council under grant 2018-03701, the European Union’s Horizon 2020 research and innovation programme under the Marie Skłodowska-Curie grant agreement H2020-MSCA-IF-2017 798063, the Spanish Ministry of
Science, Innovation and Universities under Projects TEC2017-89925-R and
PRX18/00638 and by the ICREA Academia Programme. We are also grateful to Dr.~Michalis Matthaiou for initial discussions on this work.      }     
%\HW{@ALL: please add text here.}
\balance 
\bibliography{references}
\end{document}

%% file: Figures/PEBvdistance.tex
% This file was created by matlab2tikz.
%
%The latest updates can be retrieved from
%  http://www.mathworks.com/matlabcentral/fileexchange/22022-matlab2tikz-matlab2tikz
%where you can also make suggestions and rate matlab2tikz.
%
\definecolor{mycolor1}{rgb}{0.00000,0.44700,0.74100}%
\definecolor{mycolor2}{rgb}{0.85000,0.32500,0.09800}%
\definecolor{mycolor3}{rgb}{0.92900,0.69400,0.12500}%
\definecolor{mycolor4}{rgb}{0.49400,0.18400,0.55600}%
\definecolor{mycolor5}{rgb}{0.46600,0.67400,0.18800}%
\begin{tikzpicture}[scale=1\columnwidth/10cm]

\begin{axis}[%
width=8cm,
height=5cm,
at={(1.011in,0.642in)},
scale only axis,
xmin=0,
xmax=15,
xlabel style={font=\color{white!15!black}},
xlabel={distance [m]},
ymode=log,
ymin=1e-04,
ymax=17.7187827502751,
yminorticks=true,
ymajorgrids=true,
xmajorgrids=true,
ylabel style={font=\color{white!15!black}},
ylabel={PEB [m]},
mark repeat={10},
axis background/.style={fill=white},
legend style={legend cell align=left, align=left, draw=white!15!black},
legend pos=south east
]
\addplot [color=mycolor1, line width=1.0pt, mark=*]
  table[row sep=crcr]{%
0.15	3.43609564934313e-06\\
0.224623115577889	1.01187398324746e-05\\
0.299246231155779	2.27661651348112e-05\\
0.373869346733668	4.3639521132112e-05\\
0.448492462311558	7.33901989270157e-05\\
0.523115577889447	0.000114714152411408\\
0.597738693467337	0.000171011166513303\\
0.672361809045226	0.000239831798954646\\
0.746984924623116	0.000331984446472919\\
0.821608040201005	0.000433580228643098\\
0.896231155778894	0.000566359532481849\\
0.970854271356784	0.00071514109696792\\
1.04547738693467	0.000897691353034525\\
1.12010050251256	0.00110595923848092\\
1.19472361809045	0.00133231059403754\\
1.26934673366834	0.00160405397060779\\
1.34396984924623	0.00188495811389207\\
1.41859296482412	0.00223922248521507\\
1.49321608040201	0.00258972966360195\\
1.5678391959799	0.00299813520152112\\
1.64246231155779	0.00347852809324681\\
1.71708542713568	0.00394278658245729\\
1.79170854271357	0.00450136263171926\\
1.86633165829146	0.00504494936083137\\
1.94095477386935	0.0057292210315598\\
2.01557788944724	0.00638320058310575\\
2.09020100502513	0.00702103165544903\\
2.16482412060301	0.00790138716288485\\
2.2394472361809	0.00881307751736501\\
2.31407035175879	0.00970250712171069\\
2.38869346733668	0.0105441755194416\\
2.46331658291457	0.0115485962348459\\
2.53793969849246	0.0126638182905389\\
2.61256281407035	0.013849546897376\\
2.68718592964824	0.015104844358986\\
2.76180904522613	0.0163006576806178\\
2.83643216080402	0.0177522597598087\\
2.91105527638191	0.0190279012102548\\
2.9856783919598	0.0208186668221285\\
3.06030150753769	0.0222783759415962\\
3.13492462311558	0.0239896445721206\\
3.20954773869347	0.0257608130517865\\
3.28417085427136	0.027547126531915\\
3.35879396984925	0.029498005243184\\
3.43341708542714	0.0314512382859311\\
3.50804020100502	0.0336743892483031\\
3.58266331658291	0.035787100671142\\
3.6572864321608	0.0378906509595799\\
3.73190954773869	0.0404148532259402\\
3.80653266331658	0.0428949853605931\\
3.88115577889447	0.0453020048697025\\
3.95577889447236	0.0476205465434938\\
4.03040201005025	0.0503623290175441\\
4.10502512562814	0.0532206596622354\\
4.17964824120603	0.0571709428159289\\
4.25427135678392	0.0594163533697392\\
4.32889447236181	0.0627814938877441\\
4.4035175879397	0.0663724741813151\\
4.47814070351759	0.0695732771944074\\
4.55276381909548	0.0730055476944384\\
4.62738693467337	0.076963307906393\\
4.70201005025126	0.0804847114288944\\
4.77663316582915	0.0838749484578522\\
4.85125628140704	0.0885450082924986\\
4.92587939698493	0.0920347064756818\\
5.00050251256281	0.0971952502116034\\
5.0751256281407	0.101656050346161\\
5.14974874371859	0.10581644555772\\
5.22437185929648	0.111148211007585\\
5.29899497487437	0.116512734135671\\
5.37361809045226	0.119788708382377\\
5.44824120603015	0.124593575419283\\
5.52286432160804	0.130858266862909\\
5.59748743718593	0.135843766528268\\
5.67211055276382	0.141315413086731\\
5.74673366834171	0.1479270296549\\
5.8213567839196	0.152621707200484\\
5.89597989949749	0.158988737074751\\
5.97060301507538	0.165247561914846\\
6.04522613065327	0.172059778719646\\
6.11984924623116	0.179664834084267\\
6.19447236180905	0.184993019253366\\
6.26909547738694	0.190878314195975\\
6.34371859296482	0.198870219128807\\
6.41834170854271	0.205540828835898\\
6.4929648241206	0.211685457409234\\
6.56758793969849	0.219715458615408\\
6.64221105527638	0.226008416952833\\
6.71683417085427	0.236568521367889\\
6.79145728643216	0.241613214781914\\
6.86608040201005	0.252564718962613\\
6.94070351758794	0.259464260278224\\
7.01532663316583	0.267354957661491\\
7.08994974874372	0.277474873589979\\
7.16457286432161	0.284834953690473\\
7.2391959798995	0.292607793490715\\
7.31381909547739	0.304732547691686\\
7.38844221105528	0.31225648980182\\
7.46306532663317	0.322996741116301\\
7.53768844221105	0.327753116993253\\
7.61231155778894	0.342117803703292\\
7.68693467336683	0.350056553490903\\
7.76155778894472	0.364803177223982\\
7.83618090452261	0.373083770110844\\
7.9108040201005	0.382726473462589\\
7.98542713567839	0.396248040340287\\
8.06005025125628	0.405201683819713\\
8.13467336683417	0.41627482123627\\
8.20929648241206	0.428760063469517\\
8.28391959798995	0.440942936554396\\
8.35854271356784	0.451965500220125\\
8.43316582914573	0.469440309929817\\
8.50778894472362	0.47598417613058\\
8.58241206030151	0.491063733041133\\
8.6570351758794	0.503629688084223\\
8.73165829145729	0.513670154940323\\
8.80628140703518	0.530422094466779\\
8.88090452261307	0.543985923678448\\
8.95552763819095	0.555958502746328\\
9.03015075376884	0.574995195176389\\
9.10477386934673	0.584221414455404\\
9.17939698492462	0.604488201449413\\
9.25402010050251	0.613257356368993\\
9.3286432160804	0.63422391515378\\
9.40326633165829	0.638959560253457\\
9.47788944723618	0.661003459376505\\
9.55251256281407	0.672592678637504\\
9.62713567839196	0.6959417208706\\
9.70175879396985	0.708596471292016\\
9.77638190954774	0.718050710604812\\
9.85100502512563	0.742248796238273\\
9.92562814070352	0.756670619702346\\
10.0002512562814	0.771092235014353\\
10.0748743718593	0.792626182525518\\
10.1494974874372	0.816452886582108\\
10.2241206030151	0.830257878671731\\
10.298743718593	0.842343139954701\\
10.3733668341709	0.866600711864578\\
10.4479899497487	0.876592184782742\\
10.5226130653266	0.907874243004911\\
10.5972361809045	0.920981176816246\\
10.6718592964824	0.941789236917046\\
10.7464824120603	0.958844203394958\\
10.8211055276382	0.978893659599575\\
10.8957286432161	0.998761975046171\\
10.970351758794	1.02607805250884\\
11.0449748743719	1.04220863119715\\
11.1195979899497	1.06813622340362\\
11.1942211055276	1.08984328242466\\
11.2688442211055	1.10138221584484\\
11.3434673366834	1.13157512102378\\
11.4180904522613	1.15321578976952\\
11.4927135678392	1.18235952555366\\
11.5673366834171	1.20243307078033\\
11.641959798995	1.22066394074797\\
11.7165829145729	1.25316726896221\\
11.7912060301508	1.26722470847296\\
11.8658291457286	1.29169268112636\\
11.9404522613065	1.30895109345296\\
12.0150753768844	1.33338658482894\\
12.0896984924623	1.36633910217999\\
12.1643216080402	1.40051521756916\\
12.2389447236181	1.42558918530796\\
12.313567839196	1.44439753544013\\
12.3881909547739	1.47739677976193\\
12.4628140703518	1.49878229923946\\
12.5374371859296	1.51663229626843\\
12.6120603015075	1.55355921909647\\
12.6866834170854	1.57710487067226\\
12.7613065326633	1.61654213478209\\
12.8359296482412	1.63709959412538\\
12.9105527638191	1.67120832295402\\
12.985175879397	1.68876963727278\\
13.0597989949749	1.71677991851514\\
13.1344221105528	1.76377088550059\\
13.2090452261307	1.78940100028439\\
13.2836683417085	1.81861245629083\\
13.3582914572864	1.8477816245109\\
13.4329145728643	1.87476712639349\\
13.5075376884422	1.90073715434494\\
13.5821608040201	1.95097443083386\\
13.656783919598	1.96627998160562\\
13.7314070351759	1.99972753378414\\
13.8060301507538	2.02855909787324\\
13.8806532663317	2.06306944395287\\
13.9552763819095	2.12627737434575\\
14.0298994974874	2.14275470291698\\
14.1045226130653	2.17018967216683\\
14.1791457286432	2.21221505463514\\
14.2537688442211	2.27102634197462\\
14.328391959799	2.27563977124847\\
14.4030150753769	2.30440858721242\\
14.4776381909548	2.32932128202798\\
14.5522613065327	2.38445503029307\\
14.6268844221106	2.42287891566942\\
14.7015075376884	2.46605849508861\\
14.7761306532663	2.51477504114036\\
14.8507537688442	2.54560020416209\\
14.9253768844221	2.57469917846897\\
15	2.59506622254323\\
};
\addlegendentry{\small{randomized}}

\addplot [color=mycolor2, line width=1.0pt, dashed]
  table[row sep=crcr]{%
0.15	3.05515948157519e-06\\
0.224623115577889	8.83347627785774e-06\\
0.299246231155779	1.95433145604395e-05\\
0.373869346733668	3.70832237569381e-05\\
0.448492462311558	6.35495881753608e-05\\
0.523115577889447	0.000102030430686705\\
0.597738693467337	0.000148804131209089\\
0.672361809045226	0.000203196626920648\\
0.746984924623116	0.000269387035032759\\
0.821608040201005	0.000369120366107605\\
0.896231155778894	0.000434641066397308\\
0.970854271356784	0.000550931888083801\\
1.04547738693467	0.000664386556026739\\
1.12010050251256	0.00079973765961378\\
1.19472361809045	0.000936848477652329\\
1.26934673366834	0.00110634864746297\\
1.34396984924623	0.00120832998781432\\
1.41859296482412	0.00139233406840658\\
1.49321608040201	0.00160858330689617\\
1.5678391959799	0.0017070131248004\\
1.64246231155779	0.00216132187458234\\
1.71708542713568	0.00224643012634248\\
1.79170854271357	0.00237952424812862\\
1.86633165829146	0.00265755188707124\\
1.94095477386935	0.00294790278624763\\
2.01557788944724	0.0032120921887513\\
2.09020100502513	0.00332952130939077\\
2.16482412060301	0.00380877782968138\\
2.2394472361809	0.00378136619702894\\
2.31407035175879	0.00424558522231382\\
2.38869346733668	0.0046580468012134\\
2.46331658291457	0.00472732453744676\\
2.53793969849246	0.00514733983150334\\
2.61256281407035	0.00567299448314191\\
2.68718592964824	0.00580051400400205\\
2.76180904522613	0.00615406393677588\\
2.83643216080402	0.00693742369094082\\
2.91105527638191	0.00722324608045216\\
2.9856783919598	0.00721109557305455\\
3.06030150753769	0.0073741105362053\\
3.13492462311558	0.0079244396106512\\
3.20954773869347	0.00834153161209875\\
3.28417085427136	0.00908692843815573\\
3.35879396984925	0.00926022592540579\\
3.43341708542714	0.00961950452072075\\
3.50804020100502	0.009890715863587\\
3.58266331658291	0.0105324515433703\\
3.6572864321608	0.0113331863471644\\
3.73190954773869	0.0115256510174353\\
3.80653266331658	0.0117728764713991\\
3.88115577889447	0.0118983820807554\\
3.95577889447236	0.0123017898838736\\
4.03040201005025	0.0136654558953314\\
4.10502512562814	0.013629883061281\\
4.17964824120603	0.0142223364584461\\
4.25427135678392	0.0140670123476792\\
4.32889447236181	0.015014635433066\\
4.4035175879397	0.0159763279529459\\
4.47814070351759	0.0167140836506065\\
4.55276381909548	0.0175999515927384\\
4.62738693467337	0.0170294896244159\\
4.70201005025126	0.0183277652180715\\
4.77663316582915	0.017960824358159\\
4.85125628140704	0.018639248966417\\
4.92587939698493	0.0196116563939366\\
5.00050251256281	0.0205786116762592\\
5.0751256281407	0.0210179521298941\\
5.14974874371859	0.0218865801023232\\
5.22437185929648	0.0219411605245161\\
5.29899497487437	0.0228840516725605\\
5.37361809045226	0.0241083247747042\\
5.44824120603015	0.0250132633310892\\
5.52286432160804	0.0250051735358049\\
5.59748743718593	0.0259449026444845\\
5.67211055276382	0.0257276393867653\\
5.74673366834171	0.0270340788962547\\
5.8213567839196	0.028099285268025\\
5.89597989949749	0.0289521399754025\\
5.97060301507538	0.0289153185782568\\
6.04522613065327	0.030341431720306\\
6.11984924623116	0.0310973799244838\\
6.19447236180905	0.0316116583237529\\
6.26909547738694	0.0331317244281866\\
6.34371859296482	0.0339708045183612\\
6.41834170854271	0.0354514849224843\\
6.4929648241206	0.0352867494953213\\
6.56758793969849	0.0373107080589641\\
6.64221105527638	0.0373617422017982\\
6.71683417085427	0.0375055003930885\\
6.79145728643216	0.0383446515037821\\
6.86608040201005	0.0405450531129326\\
6.94070351758794	0.040041012268693\\
7.01532663316583	0.0420766423519818\\
7.08994974874372	0.0432467342036901\\
7.16457286432161	0.0424484805237506\\
7.2391959798995	0.0438468662151086\\
7.31381909547739	0.0461207773933402\\
7.38844221105528	0.0453224675666962\\
7.46306532663317	0.0473741924870856\\
7.53768844221105	0.0493376179175808\\
7.61231155778894	0.0487159950247349\\
7.68693467336683	0.0508115345254513\\
7.76155778894472	0.0504022967334573\\
7.83618090452261	0.0532240018850189\\
7.9108040201005	0.0539320220216093\\
7.98542713567839	0.0551921029345248\\
8.06005025125628	0.0559722832600597\\
8.13467336683417	0.0570459920326499\\
8.20929648241206	0.0588213148927062\\
8.28391959798995	0.0592365497205686\\
8.35854271356784	0.060563598274278\\
8.43316582914573	0.0613186562863595\\
8.50778894472362	0.062373107673197\\
8.58241206030151	0.0645316186627577\\
8.6570351758794	0.0666656165308747\\
8.73165829145729	0.0676101965702306\\
8.80628140703518	0.067293679164174\\
8.88090452261307	0.0684073478944344\\
8.95552763819095	0.0709629092889901\\
9.03015075376884	0.0703153549894874\\
9.10477386934673	0.072289235793333\\
9.17939698492462	0.0743880237935725\\
9.25402010050251	0.0755033811389179\\
9.3286432160804	0.0765841969337083\\
9.40326633165829	0.0774503895374923\\
9.47788944723618	0.0811361765505142\\
9.55251256281407	0.0806253138848253\\
9.62713567839196	0.0828757435401411\\
9.70175879396985	0.084937764822617\\
9.77638190954774	0.0851003533338684\\
9.85100502512563	0.0858338545475136\\
9.92562814070352	0.0883372096739361\\
10.0002512562814	0.0886459933747889\\
10.0748743718593	0.0901249731524359\\
10.1494974874372	0.0932384971970457\\
10.2241206030151	0.0956719752641884\\
10.298743718593	0.0966625390882299\\
10.3733668341709	0.0983275204396851\\
10.4479899497487	0.0990658961516197\\
10.5226130653266	0.100856577562763\\
10.5972361809045	0.102026795598432\\
10.6718592964824	0.104916102406231\\
10.7464824120603	0.104865302073059\\
10.8211055276382	0.107208517040462\\
10.8957286432161	0.108844418768995\\
10.970351758794	0.10903863803465\\
11.0449748743719	0.111653586532748\\
11.1195979899497	0.112712296739639\\
11.1942211055276	0.115404511212047\\
11.2688442211055	0.115175924718559\\
11.3434673366834	0.119482948187111\\
11.4180904522613	0.120436113543196\\
11.4927135678392	0.123911863278553\\
11.5673366834171	0.123093245897292\\
11.641959798995	0.12671774578052\\
11.7165829145729	0.129024073054092\\
11.7912060301508	0.128521778127959\\
11.8658291457286	0.132959479201209\\
11.9404522613065	0.130996611602349\\
12.0150753768844	0.135768420649282\\
12.0896984924623	0.137096161962032\\
12.1643216080402	0.139012820075533\\
12.2389447236181	0.140897387949459\\
12.313567839196	0.143446212493986\\
12.3881909547739	0.144761104682361\\
12.4628140703518	0.148236656168379\\
12.5374371859296	0.147152229040068\\
12.6120603015075	0.152043246960691\\
12.6866834170854	0.153956499150921\\
12.7613065326633	0.156756102899478\\
12.8359296482412	0.158531262919901\\
12.9105527638191	0.158784383630844\\
12.985175879397	0.159567448843634\\
13.0597989949749	0.163674301854171\\
13.1344221105528	0.166762594177271\\
13.2090452261307	0.166944597425304\\
13.2836683417085	0.170973165116947\\
13.3582914572864	0.173442588133205\\
13.4329145728643	0.177137957048307\\
13.5075376884422	0.178717053448398\\
13.5821608040201	0.180092529790819\\
13.656783919598	0.184225315716944\\
13.7314070351759	0.184503842310258\\
13.8060301507538	0.187007177306641\\
13.8806532663317	0.189205022236668\\
13.9552763819095	0.191872396012051\\
14.0298994974874	0.1918235359443\\
14.1045226130653	0.19622775695115\\
14.1791457286432	0.200056822933084\\
14.2537688442211	0.20238744626912\\
14.328391959799	0.205966268979692\\
14.4030150753769	0.203690841047722\\
14.4776381909548	0.207163416368288\\
14.5522613065327	0.212229862757921\\
14.6268844221106	0.212809090360504\\
14.7015075376884	0.215133478023213\\
14.7761306532663	0.220397264386806\\
14.8507537688442	0.226572615772129\\
14.9253768844221	0.226713942469436\\
15	0.227751058411429\\
};
\addlegendentry{\small{directional, $\sigma = 1$ m}}

\addplot [color=mycolor3,line width=1.0pt, mark=+, dotted]
  table[row sep=crcr]{%
0.15	2.92729830782119e-06\\
0.224623115577889	8.67226550629602e-06\\
0.299246231155779	2.10284714855724e-05\\
0.373869346733668	3.99538540134693e-05\\
0.448492462311558	6.54112996372755e-05\\
0.523115577889447	0.000100349382725205\\
0.597738693467337	0.000165264240409323\\
0.672361809045226	0.000215823934905768\\
0.746984924623116	0.000303882063044703\\
0.821608040201005	0.000390804076276547\\
0.896231155778894	0.000475968370854016\\
0.970854271356784	0.000603299724763901\\
1.04547738693467	0.000674118743799156\\
1.12010050251256	0.000908062542256128\\
1.19472361809045	0.00097128926773712\\
1.26934673366834	0.00122547003794419\\
1.34396984924623	0.001314090924813\\
1.41859296482412	0.00157069710529968\\
1.49321608040201	0.00176377528510987\\
1.5678391959799	0.00175046537154939\\
1.64246231155779	0.00207337117061348\\
1.71708542713568	0.00238001781029358\\
1.79170854271357	0.00261367887541365\\
1.86633165829146	0.0031620247031714\\
1.94095477386935	0.00319112363395828\\
2.01557788944724	0.00349102392088584\\
2.09020100502513	0.00340691449201404\\
2.16482412060301	0.00371617917786257\\
2.2394472361809	0.00387860041389785\\
2.31407035175879	0.00455980031737768\\
2.38869346733668	0.00489807095473568\\
2.46331658291457	0.00506684772998796\\
2.53793969849246	0.0056045947853943\\
2.61256281407035	0.005688428246943\\
2.68718592964824	0.00583257712135903\\
2.76180904522613	0.00579937344362495\\
2.83643216080402	0.00644555599160182\\
2.91105527638191	0.00719147044855589\\
2.9856783919598	0.00766504754813014\\
3.06030150753769	0.00784786369441232\\
3.13492462311558	0.00782076378330924\\
3.20954773869347	0.00905380528554479\\
3.28417085427136	0.00903720379318958\\
3.35879396984925	0.00908569014365247\\
3.43341708542714	0.00936934468380769\\
3.50804020100502	0.00980903004432632\\
3.58266331658291	0.0102327179543091\\
3.6572864321608	0.0104420063345116\\
3.73190954773869	0.0113158081826834\\
3.80653266331658	0.012033914345023\\
3.88115577889447	0.012686765173785\\
3.95577889447236	0.0126370162116495\\
4.03040201005025	0.0133717764940828\\
4.10502512562814	0.0137620466716142\\
4.17964824120603	0.0143122378686737\\
4.25427135678392	0.0146293185572588\\
4.32889447236181	0.015783922704128\\
4.4035175879397	0.0157218714645952\\
4.47814070351759	0.0169573732873739\\
4.55276381909548	0.017228913270819\\
4.62738693467337	0.0174112045505463\\
4.70201005025126	0.0174653633631807\\
4.77663316582915	0.0187356171990012\\
4.85125628140704	0.019160448199764\\
4.92587939698493	0.0190915871991368\\
5.00050251256281	0.0210161245903355\\
5.0751256281407	0.0207455519321984\\
5.14974874371859	0.0212758829876894\\
5.22437185929648	0.0220225067159968\\
5.29899497487437	0.0221430202394698\\
5.37361809045226	0.0240390993866177\\
5.44824120603015	0.0241410466316474\\
5.52286432160804	0.025618336825105\\
5.59748743718593	0.0256065621152738\\
5.67211055276382	0.0262823346386752\\
5.74673366834171	0.0274086728882669\\
5.8213567839196	0.0274464646224522\\
5.89597989949749	0.0290331912215632\\
5.97060301507538	0.029381476831837\\
6.04522613065327	0.0300407805374135\\
6.11984924623116	0.0315818032596971\\
6.19447236180905	0.032723202555777\\
6.26909547738694	0.0321384800385185\\
6.34371859296482	0.0336142827026403\\
6.41834170854271	0.0347770279606703\\
6.4929648241206	0.0346145302023053\\
6.56758793969849	0.0349494654484264\\
6.64221105527638	0.0371951690536262\\
6.71683417085427	0.0371569891274315\\
6.79145728643216	0.0389059243630508\\
6.86608040201005	0.0401259132211456\\
6.94070351758794	0.0411573486533651\\
7.01532663316583	0.0404146207919241\\
7.08994974874372	0.0420579938753218\\
7.16457286432161	0.0441016821460417\\
7.2391959798995	0.0447042156361021\\
7.31381909547739	0.0471018039977505\\
7.38844221105528	0.0460509228495376\\
7.46306532663317	0.0477359721022037\\
7.53768844221105	0.0474966387230186\\
7.61231155778894	0.0483301660834551\\
7.68693467336683	0.0512989593973064\\
7.76155778894472	0.0513792553642557\\
7.83618090452261	0.0526466379948107\\
7.9108040201005	0.0521714296909307\\
7.98542713567839	0.054639326308482\\
8.06005025125628	0.0558217717956223\\
8.13467336683417	0.0562349189352821\\
8.20929648241206	0.0565880993195533\\
8.28391959798995	0.0593460549322637\\
8.35854271356784	0.059694143545938\\
8.43316582914573	0.0600927250655517\\
8.50778894472362	0.0626745372031883\\
8.58241206030151	0.0647722263066059\\
8.6570351758794	0.0650762262718399\\
8.73165829145729	0.065768349702159\\
8.80628140703518	0.0677466421729361\\
8.88090452261307	0.0690603616203253\\
8.95552763819095	0.0698217984114806\\
9.03015075376884	0.0718612688130658\\
9.10477386934673	0.0715315429996725\\
9.17939698492462	0.0750150111338459\\
9.25402010050251	0.0746628829755286\\
9.3286432160804	0.0772280437016053\\
9.40326633165829	0.0772501264965231\\
9.47788944723618	0.0775698111892227\\
9.55251256281407	0.0816180274668263\\
9.62713567839196	0.0809202127029437\\
9.70175879396985	0.0854798399319226\\
9.77638190954774	0.0864739596637923\\
9.85100502512563	0.0863063132818522\\
9.92562814070352	0.0878305956249999\\
10.0002512562814	0.0889152493797231\\
10.0748743718593	0.0924157637338459\\
10.1494974874372	0.0915512083220108\\
10.2241206030151	0.0950521802500098\\
10.298743718593	0.0956669112293774\\
10.3733668341709	0.095508031500858\\
10.4479899497487	0.0976709452620292\\
10.5226130653266	0.0990191083377453\\
10.5972361809045	0.102163936368106\\
10.6718592964824	0.101341635153946\\
10.7464824120603	0.10415390180119\\
10.8211055276382	0.105430817474904\\
10.8957286432161	0.10729008003097\\
10.970351758794	0.109345724954183\\
11.0449748743719	0.111806656650637\\
11.1195979899497	0.113313326200021\\
11.1942211055276	0.113665396238867\\
11.2688442211055	0.118631732788305\\
11.3434673366834	0.118360483224745\\
11.4180904522613	0.119448288847621\\
11.4927135678392	0.122787105324486\\
11.5673366834171	0.122401734264483\\
11.641959798995	0.1257727478513\\
11.7165829145729	0.127685233891203\\
11.7912060301508	0.127839620353893\\
11.8658291457286	0.129517436118558\\
11.9404522613065	0.1322938659238\\
12.0150753768844	0.134097423568627\\
12.0896984924623	0.135345036638981\\
12.1643216080402	0.138351496782388\\
12.2389447236181	0.139875080763086\\
12.313567839196	0.142331412339017\\
12.3881909547739	0.143510811441933\\
12.4628140703518	0.14637296545593\\
12.5374371859296	0.148506560598777\\
12.6120603015075	0.15059427668234\\
12.6866834170854	0.153435705774593\\
12.7613065326633	0.155913282768794\\
12.8359296482412	0.158544812240792\\
12.9105527638191	0.157969837151058\\
12.985175879397	0.162139226122483\\
13.0597989949749	0.164099230571137\\
13.1344221105528	0.167090589243673\\
13.2090452261307	0.166743173629542\\
13.2836683417085	0.173202329749708\\
13.3582914572864	0.172450729601012\\
13.4329145728643	0.175900735241288\\
13.5075376884422	0.178116188186454\\
13.5821608040201	0.180064818036398\\
13.656783919598	0.182783211986778\\
13.7314070351759	0.183195726382116\\
13.8060301507538	0.186322891367374\\
13.8806532663317	0.186714567007066\\
13.9552763819095	0.193102952840449\\
14.0298994974874	0.193249607526942\\
14.1045226130653	0.198489863403892\\
14.1791457286432	0.200177855999773\\
14.2537688442211	0.201563641214926\\
14.328391959799	0.205922029741232\\
14.4030150753769	0.205755021248486\\
14.4776381909548	0.208136112197584\\
14.5522613065327	0.209914359070373\\
14.6268844221106	0.218200444018603\\
14.7015075376884	0.216942717907215\\
14.7761306532663	0.216529545998041\\
14.8507537688442	0.222536420720392\\
14.9253768844221	0.224989242796191\\
15	0.227279840756451\\
};
\addlegendentry{\small{positional, $\sigma = 1$ m}}

\addplot [color=mycolor4,line width=1.0pt, mark=o]
  table[row sep=crcr]{%
0.15	2.27036582171447e-06\\
0.224623115577889	5.21176102679154e-06\\
0.299246231155779	9.30532630275853e-06\\
0.373869346733668	1.45040710390254e-05\\
0.448492462311558	2.11127743931183e-05\\
0.523115577889447	2.89645424329274e-05\\
0.597738693467337	3.79120659180691e-05\\
0.672361809045226	4.85355078358684e-05\\
0.746984924623116	5.96225320454041e-05\\
0.821608040201005	7.38741560009398e-05\\
0.896231155778894	8.92003272635609e-05\\
0.970854271356784	0.000105524979745956\\
1.04547738693467	0.000124313807403344\\
1.12010050251256	0.000142857183799016\\
1.19472361809045	0.00016587835888811\\
1.26934673366834	0.000189531679334522\\
1.34396984924623	0.000215710870886113\\
1.41859296482412	0.00024828424762435\\
1.49321608040201	0.000275126586212191\\
1.5678391959799	0.000308858057895827\\
1.64246231155779	0.000345080947737134\\
1.71708542713568	0.000386669949504552\\
1.79170854271357	0.00042718275140285\\
1.86633165829146	0.000471622131086052\\
1.94095477386935	0.000521675928991621\\
2.01557788944724	0.000569914828203497\\
2.09020100502513	0.000626323794699997\\
2.16482412060301	0.000682458020564578\\
2.2394472361809	0.000747686632488112\\
2.31407035175879	0.000816866947914559\\
2.38869346733668	0.000888606069056426\\
2.46331658291457	0.000954272137460246\\
2.53793969849246	0.00104041760227116\\
2.61256281407035	0.00112862929406036\\
2.68718592964824	0.00122733782719806\\
2.76180904522613	0.00132557475524672\\
2.83643216080402	0.00142900508423924\\
2.91105527638191	0.00152940094160996\\
2.9856783919598	0.00166082874142332\\
3.06030150753769	0.00177296208113536\\
3.13492462311558	0.00189253814026853\\
3.20954773869347	0.00206335446264191\\
3.28417085427136	0.0021853940257933\\
3.35879396984925	0.00232557709356482\\
3.43341708542714	0.00250269001747485\\
3.50804020100502	0.00268163044453731\\
3.58266331658291	0.00286831127837019\\
3.6572864321608	0.00304273885517605\\
3.73190954773869	0.0032265192608975\\
3.80653266331658	0.00347934944202106\\
3.88115577889447	0.00369947142807089\\
3.95577889447236	0.0039430489451965\\
4.03040201005025	0.00417183936120616\\
4.10502512562814	0.00435980243751039\\
4.17964824120603	0.00467316243282239\\
4.25427135678392	0.00498481424565317\\
4.32889447236181	0.00524893685740264\\
4.4035175879397	0.00554909284586462\\
4.47814070351759	0.0059183399887166\\
4.55276381909548	0.00633452051794757\\
4.62738693467337	0.00663208071407284\\
4.70201005025126	0.00700329581860007\\
4.77663316582915	0.00737597715409008\\
4.85125628140704	0.00778987833054267\\
4.92587939698493	0.00814525349477839\\
5.00050251256281	0.00876462397625155\\
5.0751256281407	0.009300450816378\\
5.14974874371859	0.00966401944889782\\
5.22437185929648	0.0103234497558755\\
5.29899497487437	0.0109041539129429\\
5.37361809045226	0.0114455544396756\\
5.44824120603015	0.0119219933130656\\
5.52286432160804	0.0125889187473314\\
5.59748743718593	0.0133172181962287\\
5.67211055276382	0.0137224112881324\\
5.74673366834171	0.0145402063574599\\
5.8213567839196	0.0152041491765944\\
5.89597989949749	0.016275959037164\\
5.97060301507538	0.0168558034487542\\
6.04522613065327	0.0180007585003295\\
6.11984924623116	0.0184552750772453\\
6.19447236180905	0.0194322388037668\\
6.26909547738694	0.0208020606664064\\
6.34371859296482	0.0214596759529261\\
6.41834170854271	0.0223937146779242\\
6.4929648241206	0.0234582956964516\\
6.56758793969849	0.024847799188232\\
6.64221105527638	0.0257247065104415\\
6.71683417085427	0.0270294509653368\\
6.79145728643216	0.0284649140792786\\
6.86608040201005	0.0296239162528982\\
6.94070351758794	0.0309476070093868\\
7.01532663316583	0.0325581264045806\\
7.08994974874372	0.0335123210335535\\
7.16457286432161	0.0351081803891416\\
7.2391959798995	0.0367352308004739\\
7.31381909547739	0.0387317071840414\\
7.38844221105528	0.0399067364769985\\
7.46306532663317	0.041864826922754\\
7.53768844221105	0.0431294496765691\\
7.61231155778894	0.0455365066102981\\
7.68693467336683	0.0469465701246395\\
7.76155778894472	0.0492313350519569\\
7.83618090452261	0.0507988837998201\\
7.9108040201005	0.0525764344101631\\
7.98542713567839	0.0555411203148597\\
8.06005025125628	0.0583813104493452\\
8.13467336683417	0.0597637150413059\\
8.20929648241206	0.0625959001521062\\
8.28391959798995	0.064559363555867\\
8.35854271356784	0.067578735114511\\
8.43316582914573	0.0719193544246922\\
8.50778894472362	0.0727480063810916\\
8.58241206030151	0.0760812749838786\\
8.6570351758794	0.0774837387217939\\
8.73165829145729	0.0821972640825564\\
8.80628140703518	0.0864200933053929\\
8.88090452261307	0.0875749601968572\\
8.95552763819095	0.0900641954911587\\
9.03015075376884	0.0947078258232864\\
9.10477386934673	0.0984339749819278\\
9.17939698492462	0.101739674834191\\
9.25402010050251	0.105734592146064\\
9.3286432160804	0.1108121448743\\
9.40326633165829	0.112810131295329\\
9.47788944723618	0.115151598329433\\
9.55251256281407	0.120985830534775\\
9.62713567839196	0.126156754653808\\
9.70175879396985	0.130246738068981\\
9.77638190954774	0.135841173529601\\
9.85100502512563	0.138494547485637\\
9.92562814070352	0.145282866480466\\
10.0002512562814	0.150017954130979\\
10.0748743718593	0.15457539110984\\
10.1494974874372	0.160532177709067\\
10.2241206030151	0.165598656112823\\
10.298743718593	0.171572089602679\\
10.3733668341709	0.175001105190825\\
10.4479899497487	0.182099777163797\\
10.5226130653266	0.186557206329735\\
10.5972361809045	0.195116555286061\\
10.6718592964824	0.202372096220627\\
10.7464824120603	0.206034796666501\\
10.8211055276382	0.213017178929238\\
10.8957286432161	0.217690518134771\\
10.970351758794	0.224981236392591\\
11.0449748743719	0.235454165718714\\
11.1195979899497	0.243629806654276\\
11.1942211055276	0.25319313352864\\
11.2688442211055	0.253682905331451\\
11.3434673366834	0.26309793435533\\
11.4180904522613	0.271782502467241\\
11.4927135678392	0.284014886744065\\
11.5673366834171	0.28933467127487\\
11.641959798995	0.300611288220883\\
11.7165829145729	0.308504069442841\\
11.7912060301508	0.315276151270348\\
11.8658291457286	0.322038403388165\\
11.9404522613065	0.333979787727655\\
12.0150753768844	0.348620824627741\\
12.0896984924623	0.353324393361479\\
12.1643216080402	0.368244794266928\\
12.2389447236181	0.37300086851993\\
12.313567839196	0.38900480174254\\
12.3881909547739	0.401454549872255\\
12.4628140703518	0.411429809301189\\
12.5374371859296	0.434925660983627\\
12.6120603015075	0.435370423649403\\
12.6866834170854	0.447026707637505\\
12.7613065326633	0.453087926348504\\
12.8359296482412	0.471436635811629\\
12.9105527638191	0.479446012734121\\
12.985175879397	0.497394085066794\\
13.0597989949749	0.51450530963262\\
13.1344221105528	0.538341200985628\\
13.2090452261307	0.538164613298217\\
13.2836683417085	0.548577066110924\\
13.3582914572864	0.581402777567107\\
13.4329145728643	0.583941599749408\\
13.5075376884422	0.610358557106037\\
13.5821608040201	0.611554815545266\\
13.656783919598	0.61612736899152\\
13.7314070351759	0.654258891757681\\
13.8060301507538	0.662195830689962\\
13.8806532663317	0.678540904684322\\
13.9552763819095	0.699025682282208\\
14.0298994974874	0.706953351163918\\
14.1045226130653	0.734897229529005\\
14.1791457286432	0.750423683838357\\
14.2537688442211	0.774662864288393\\
14.328391959799	0.792799086166622\\
14.4030150753769	0.827146452865954\\
14.4776381909548	0.838097527728124\\
14.5522613065327	0.863605574778675\\
14.6268844221106	0.881915462039738\\
14.7015075376884	0.894512450632799\\
14.7761306532663	0.922067063491167\\
14.8507537688442	0.934789854068227\\
14.9253768844221	0.967444648443281\\
15	0.980934481152936\\
};
\addlegendentry{\small{directional, $\sigma = 0.1$ m}}

\addplot [color=mycolor5,line width=1.0pt, mark=x]
  table[row sep=crcr]{%
0.15	1.63465996683556e-06\\
0.224623115577889	4.0360777863755e-06\\
0.299246231155779	7.28718544148081e-06\\
0.373869346733668	1.14045500742025e-05\\
0.448492462311558	1.66716301207586e-05\\
0.523115577889447	2.2787311452934e-05\\
0.597738693467337	3.03141391050034e-05\\
0.672361809045226	3.85007904911743e-05\\
0.746984924623116	4.8231022271079e-05\\
0.821608040201005	5.83737752131512e-05\\
0.896231155778894	7.11891831303911e-05\\
0.970854271356784	8.44099691349461e-05\\
1.04547738693467	9.91338233051662e-05\\
1.12010050251256	0.000117522857181597\\
1.19472361809045	0.000135281278397446\\
1.26934673366834	0.000153124382906627\\
1.34396984924623	0.000177393277088284\\
1.41859296482412	0.000200421252463596\\
1.49321608040201	0.000228964997401528\\
1.5678391959799	0.000256259563099306\\
1.64246231155779	0.000284993365079854\\
1.71708542713568	0.000318033482171532\\
1.79170854271357	0.000354179943226142\\
1.86633165829146	0.000391083216964658\\
1.94095477386935	0.000433720646794762\\
2.01557788944724	0.000478848626850638\\
2.09020100502513	0.000526557928739233\\
2.16482412060301	0.000575974128483531\\
2.2394472361809	0.000634962943325645\\
2.31407035175879	0.000693086749938243\\
2.38869346733668	0.00074972050726609\\
2.46331658291457	0.000817829223016328\\
2.53793969849246	0.00089998960525724\\
2.61256281407035	0.00096565701616676\\
2.68718592964824	0.00106092822740793\\
2.76180904522613	0.00113763497294737\\
2.83643216080402	0.0012256681203167\\
2.91105527638191	0.00133350500350824\\
2.9856783919598	0.00144892069179271\\
3.06030150753769	0.00156111189441593\\
3.13492462311558	0.00168054587558379\\
3.20954773869347	0.00181354588617495\\
3.28417085427136	0.00194003518062955\\
3.35879396984925	0.00207394215423447\\
3.43341708542714	0.00221695674680282\\
3.50804020100502	0.00237797434438179\\
3.58266331658291	0.00254857316333281\\
3.6572864321608	0.00272976192213334\\
3.73190954773869	0.00291375179006475\\
3.80653266331658	0.00312997776269582\\
3.88115577889447	0.00334435692360704\\
3.95577889447236	0.00356421735271523\\
4.03040201005025	0.00379782514642626\\
4.10502512562814	0.00402627501346371\\
4.17964824120603	0.00427167012336866\\
4.25427135678392	0.00456084815885704\\
4.32889447236181	0.00485866068004796\\
4.4035175879397	0.00514079778312156\\
4.47814070351759	0.0055016757690444\\
4.55276381909548	0.005841497364566\\
4.62738693467337	0.00610977661098963\\
4.70201005025126	0.00652939403997634\\
4.77663316582915	0.00686495209515534\\
4.85125628140704	0.00723411193528074\\
4.92587939698493	0.00782005718326527\\
5.00050251256281	0.00817799403695449\\
5.0751256281407	0.00868833703967736\\
5.14974874371859	0.00902998630070438\\
5.22437185929648	0.00958568830989762\\
5.29899497487437	0.0100434257760961\\
5.37361809045226	0.0105715633428768\\
5.44824120603015	0.0112544460078429\\
5.52286432160804	0.0116683441310681\\
5.59748743718593	0.0125053872643298\\
5.67211055276382	0.0131943936805937\\
5.74673366834171	0.013845177022395\\
5.8213567839196	0.014597045779349\\
5.89597989949749	0.015419482253381\\
5.97060301507538	0.0159565430135999\\
6.04522613065327	0.0169439082119913\\
6.11984924623116	0.0178678451042447\\
6.19447236180905	0.0187988430889419\\
6.26909547738694	0.0195766284639162\\
6.34371859296482	0.0206631968293644\\
6.41834170854271	0.0215723212673752\\
6.4929648241206	0.0223759930123709\\
6.56758793969849	0.0235415815124179\\
6.64221105527638	0.0248483214745481\\
6.71683417085427	0.0263455975945499\\
6.79145728643216	0.0271862831403753\\
6.86608040201005	0.0282469495427635\\
6.94070351758794	0.0294313504764146\\
7.01532663316583	0.0303621599526782\\
7.08994974874372	0.0321859231245167\\
7.16457286432161	0.0340462226719917\\
7.2391959798995	0.0354490700086509\\
7.31381909547739	0.036584483353127\\
7.38844221105528	0.039029905042607\\
7.46306532663317	0.039749685092337\\
7.53768844221105	0.0416870255316476\\
7.61231155778894	0.0442944027490449\\
7.68693467336683	0.0455458116403791\\
7.76155778894472	0.047251434404086\\
7.83618090452261	0.0499730070624818\\
7.9108040201005	0.0524896241465561\\
7.98542713567839	0.0538108223718744\\
8.06005025125628	0.055863137694063\\
8.13467336683417	0.0590113935320871\\
8.20929648241206	0.0619855348536536\\
8.28391959798995	0.0644698957414246\\
8.35854271356784	0.0665565686106392\\
8.43316582914573	0.0669717624148897\\
8.50778894472362	0.0698107278286885\\
8.58241206030151	0.0743753915718349\\
8.6570351758794	0.0775967755690539\\
8.73165829145729	0.0795611810582051\\
8.80628140703518	0.0821153957023816\\
8.88090452261307	0.0852763311508144\\
8.95552763819095	0.0903113350370442\\
9.03015075376884	0.0925022021855198\\
9.10477386934673	0.0969096831551155\\
9.17939698492462	0.100994792921413\\
9.25402010050251	0.105221708454999\\
9.3286432160804	0.107687588305373\\
9.40326633165829	0.111290904012158\\
9.47788944723618	0.114346913612875\\
9.55251256281407	0.117910974980048\\
9.62713567839196	0.124710608359997\\
9.70175879396985	0.127013691345975\\
9.77638190954774	0.13369159460983\\
9.85100502512563	0.14036415904466\\
9.92562814070352	0.141708837091006\\
10.0002512562814	0.148313141677229\\
10.0748743718593	0.152712917519598\\
10.1494974874372	0.158319120052328\\
10.2241206030151	0.161371626818373\\
10.298743718593	0.169704635938623\\
10.3733668341709	0.170991988050941\\
10.4479899497487	0.17534002056637\\
10.5226130653266	0.184376511926182\\
10.5972361809045	0.192264451071098\\
10.6718592964824	0.193992745266942\\
10.7464824120603	0.207586044688355\\
10.8211055276382	0.210786228046097\\
10.8957286432161	0.221423000799703\\
10.970351758794	0.220981639612588\\
11.0449748743719	0.234648257714148\\
11.1195979899497	0.236069360072089\\
11.1942211055276	0.242562710461724\\
11.2688442211055	0.251452827141721\\
11.3434673366834	0.255239063823043\\
11.4180904522613	0.265055377568618\\
11.4927135678392	0.274589959700427\\
11.5673366834171	0.293414723320298\\
11.641959798995	0.294787049285963\\
11.7165829145729	0.300668369732612\\
11.7912060301508	0.306169729191348\\
11.8658291457286	0.319768017617929\\
11.9404522613065	0.322575179957183\\
12.0150753768844	0.337046280847428\\
12.0896984924623	0.350367446816934\\
12.1643216080402	0.367755805091001\\
12.2389447236181	0.366010326395906\\
12.313567839196	0.385262574143821\\
12.3881909547739	0.39425845490057\\
12.4628140703518	0.402806963057582\\
12.5374371859296	0.416228009600066\\
12.6120603015075	0.423726002775024\\
12.6866834170854	0.4333024915754\\
12.7613065326633	0.460062027451153\\
12.8359296482412	0.4742884595827\\
12.9105527638191	0.460888240059692\\
12.985175879397	0.502000484722851\\
13.0597989949749	0.520064672856325\\
13.1344221105528	0.51597862163484\\
13.2090452261307	0.543101152362861\\
13.2836683417085	0.564315736214553\\
13.3582914572864	0.558621638122086\\
13.4329145728643	0.574818679446917\\
13.5075376884422	0.587868536839689\\
13.5821608040201	0.599203822238086\\
13.656783919598	0.610992779318275\\
13.7314070351759	0.644499512135668\\
13.8060301507538	0.666076401951283\\
13.8806532663317	0.685097049570176\\
13.9552763819095	0.709172385194516\\
14.0298994974874	0.701272556005677\\
14.1045226130653	0.73164985945223\\
14.1791457286432	0.740680530732458\\
14.2537688442211	0.775114666615415\\
14.328391959799	0.780741865036898\\
14.4030150753769	0.809585541837997\\
14.4776381909548	0.816441197932174\\
14.5522613065327	0.856035407911709\\
14.6268844221106	0.868161228976665\\
14.7015075376884	0.876733882472387\\
14.7761306532663	0.931846212857264\\
14.8507537688442	0.906976165005444\\
14.9253768844221	0.948406497172498\\
15	0.98770513018585\\
};
\addlegendentry{\small{positional, $\sigma = 0.1$ m}}
\addplot[mark=none, black, samples=10] coordinates {(0,1.73) (15,1.73)};
\node[above] at (2.6,1.5) {\small{prior PEB $\sigma = 1$ m}};
\addplot[mark=none, black, samples=10,dashed] coordinates {(0,0.173) (15,0.173)};
\node[above] at (2.8,0.15) {\small{prior PEB $\sigma = 0.1$ m}};

\end{axis}
\end{tikzpicture}%

%% file: Figures/RMSEvdistance.tex
% This file was created by matlab2tikz.
%
%The latest updates can be retrieved from
%  http://www.mathworks.com/matlabcentral/fileexchange/22022-matlab2tikz-matlab2tikz
%where you can also make suggestions and rate matlab2tikz.
%
\definecolor{mycolor1}{rgb}{0.00000,0.44700,0.74100}%
\definecolor{mycolor2}{rgb}{0.49400,0.18400,0.55600}%
\definecolor{mycolor3}{rgb}{0.46600,0.67400,0.18800}
\begin{tikzpicture}[scale=1\columnwidth/10cm]

\begin{axis}[%
width=8cm,
height=5cm,
at={(1.011in,0.642in)},
scale only axis,
xmin=0,
xmax=15,
xlabel style={font=\color{white!15!black}},
xlabel={distance [m]},
ymode=log,
ymin=0.0001,
ymax=17.7187827502751,
yminorticks=true,
ymajorgrids=true,
xmajorgrids=true,
ylabel style={font=\color{white!15!black}},
ylabel={RMSE [m]},
axis background/.style={fill=white},
legend style={legend cell align=left, align=left, draw=white!15!black},
legend pos=south east
]

\addplot [color=mycolor2,line width=1.0pt, mark=o]
  table[row sep=crcr]{%
%0.15	0.507886109008708\\
%0.688888888888889	0.546695906762061\\
%1.22777777777778	0.463907344049464\\
%1.76666666666667	0.704003150504618\\
%2.30555555555556	0.889251176819501\\
%2.84444444444444	1.78504993579503\\
%3.38333333333333	2.78686463509951\\
%3.92222222222222	4.34185174913177\\
%4.46111111111111	5.28174451238512\\
%5	6.82794573902899\\
%5.5	7.18299813807544\\
%7.875	12.3335015410465\\
%10.25	14.3780100996528\\
%12.625	15.255228554967\\
%15	16.1813823820241\\
0.15	0.131732984870111\\
1	0.253525733044354\\
1.5	0.354438651034034\\
2	0.586040539688626\\
3	0.990537189384664\\
4	2.78602089564626\\
6	8.1352360058964\\
8	10.7769849740883\\
10	11.4689467891779\\
12	14.7828888753571\\
15	15.5057942740986\\
};
\addlegendentry{\small{directional}}%, $\sigma = 0.1$ m, $3 \times 1\mathrm{D}$}}

\addplot [color=mycolor3,line width=1.0pt, mark=*]
  table[row sep=crcr]{%
%0.15	0.0338358785998036\\
%0.405263157894737	0.0271550997733176\\
%0.660526315789474	0.0243530398503693\\
%0.91578947368421	0.0247026879003379\\
%1.17105263157895	0.0325641675918275\\
%1.42631578947368	0.0292364427251157\\
%1.68157894736842	0.0431548160123116\\
%1.93684210526316	0.0360844600941704\\
%2.19210526315789	0.0669166633603238\\
%2.44736842105263	0.0572604473671831\\
%2.70263157894737	0.0627831115918638\\
%2.9578947368421	0.0823699289334326\\
%3.21315789473684	0.0745397318738009\\
%3.46842105263158	0.107229535820163\\
%3.72368421052632	0.125065976738482\\
%3.97894736842105	0.154587430055977\\
%4.23421052631579	0.228985602478798\\
%4.48947368421053	0.24449628144311\\
%4.74473684210526	0.255297453930825\\
%5	0.248477640384765\\
%5.5	0.261121430471261\\
%6.55555555555556	0.439945423385246\\%
%7.61111111111111	0.526899669838466\\
%8.66666666666667	0.764697566248363\\
%9.72222222222222	0.9040\\
%10.7777777777778	1.1992\\
%11.8333333333333	1.8960\\
%12.8888888888889	1.8086\\
%13.9444444444444	2.6389\\
%15	2.6095\\
0.15	0.0101\\
1	0.0210318846713514\\
1.5	0.0295371535208284\\
2	0.0367388431085355\\
3	0.0541345837011121\\
4	0.102997228894953\\
6	0.262335957289047\\
8	0.527859056724871\\
10	0.879044312500068\\
12	1.36989574042624\\
15	2.34028286627046\\
};
\addlegendentry{\small{randomized}}%, $3 \times 1\mathrm{D}$}}

%\addplot [color=mycolor1, mark=+]
%  table[row sep=crcr]{%
%0.15	0.0272704606937009\\
%0.15 0.0123\\
%1	0.0131215424087861\\
%1 0.0154\\
%1.5	0.0112099536681165\\
%1.5 0.0124\\
%2 0.0193\\
%3	0.0307757388288759\\
%4	0.0627329524768199\\
%6	0.166178085145248\\
%8	0.371571137006021\\
%10	0.693794328513537\\
%12	1.19269143030544\\
%15	2.03387571402396\\
%};

%\addlegendentry{\small{randomized, $2\mathrm{D}+1\mathrm{D}$}}

\addplot [color=mycolor1, line width=1.0pt,dotted]
  table[row sep=crcr]{%
0.15	3.43609564934313e-06\\
0.224623115577889	1.01187398324746e-05\\
0.299246231155779	2.27661651348112e-05\\
0.373869346733668	4.3639521132112e-05\\
0.448492462311558	7.33901989270157e-05\\
0.523115577889447	0.000114714152411408\\
0.597738693467337	0.000171011166513303\\
0.672361809045226	0.000239831798954646\\
0.746984924623116	0.000331984446472919\\
0.821608040201005	0.000433580228643098\\
0.896231155778894	0.000566359532481849\\
0.970854271356784	0.00071514109696792\\
1.04547738693467	0.000897691353034525\\
1.12010050251256	0.00110595923848092\\
1.19472361809045	0.00133231059403754\\
1.26934673366834	0.00160405397060779\\
1.34396984924623	0.00188495811389207\\
1.41859296482412	0.00223922248521507\\
1.49321608040201	0.00258972966360195\\
1.5678391959799	0.00299813520152112\\
1.64246231155779	0.00347852809324681\\
1.71708542713568	0.00394278658245729\\
1.79170854271357	0.00450136263171926\\
1.86633165829146	0.00504494936083137\\
1.94095477386935	0.0057292210315598\\
2.01557788944724	0.00638320058310575\\
2.09020100502513	0.00702103165544903\\
2.16482412060301	0.00790138716288485\\
2.2394472361809	0.00881307751736501\\
2.31407035175879	0.00970250712171069\\
2.38869346733668	0.0105441755194416\\
2.46331658291457	0.0115485962348459\\
2.53793969849246	0.0126638182905389\\
2.61256281407035	0.013849546897376\\
2.68718592964824	0.015104844358986\\
2.76180904522613	0.0163006576806178\\
2.83643216080402	0.0177522597598087\\
2.91105527638191	0.0190279012102548\\
2.9856783919598	0.0208186668221285\\
3.06030150753769	0.0222783759415962\\
3.13492462311558	0.0239896445721206\\
3.20954773869347	0.0257608130517865\\
3.28417085427136	0.027547126531915\\
3.35879396984925	0.029498005243184\\
3.43341708542714	0.0314512382859311\\
3.50804020100502	0.0336743892483031\\
3.58266331658291	0.035787100671142\\
3.6572864321608	0.0378906509595799\\
3.73190954773869	0.0404148532259402\\
3.80653266331658	0.0428949853605931\\
3.88115577889447	0.0453020048697025\\
3.95577889447236	0.0476205465434938\\
4.03040201005025	0.0503623290175441\\
4.10502512562814	0.0532206596622354\\
4.17964824120603	0.0571709428159289\\
4.25427135678392	0.0594163533697392\\
4.32889447236181	0.0627814938877441\\
4.4035175879397	0.0663724741813151\\
4.47814070351759	0.0695732771944074\\
4.55276381909548	0.0730055476944384\\
4.62738693467337	0.076963307906393\\
4.70201005025126	0.0804847114288944\\
4.77663316582915	0.0838749484578522\\
4.85125628140704	0.0885450082924986\\
4.92587939698493	0.0920347064756818\\
5.00050251256281	0.0971952502116034\\
5.0751256281407	0.101656050346161\\
5.14974874371859	0.10581644555772\\
5.22437185929648	0.111148211007585\\
5.29899497487437	0.116512734135671\\
5.37361809045226	0.119788708382377\\
5.44824120603015	0.124593575419283\\
5.52286432160804	0.130858266862909\\
5.59748743718593	0.135843766528268\\
5.67211055276382	0.141315413086731\\
5.74673366834171	0.1479270296549\\
5.8213567839196	0.152621707200484\\
5.89597989949749	0.158988737074751\\
5.97060301507538	0.165247561914846\\
6.04522613065327	0.172059778719646\\
6.11984924623116	0.179664834084267\\
6.19447236180905	0.184993019253366\\
6.26909547738694	0.190878314195975\\
6.34371859296482	0.198870219128807\\
6.41834170854271	0.205540828835898\\
6.4929648241206	0.211685457409234\\
6.56758793969849	0.219715458615408\\
6.64221105527638	0.226008416952833\\
6.71683417085427	0.236568521367889\\
6.79145728643216	0.241613214781914\\
6.86608040201005	0.252564718962613\\
6.94070351758794	0.259464260278224\\
7.01532663316583	0.267354957661491\\
7.08994974874372	0.277474873589979\\
7.16457286432161	0.284834953690473\\
7.2391959798995	0.292607793490715\\
7.31381909547739	0.304732547691686\\
7.38844221105528	0.31225648980182\\
7.46306532663317	0.322996741116301\\
7.53768844221105	0.327753116993253\\
7.61231155778894	0.342117803703292\\
7.68693467336683	0.350056553490903\\
7.76155778894472	0.364803177223982\\
7.83618090452261	0.373083770110844\\
7.9108040201005	0.382726473462589\\
7.98542713567839	0.396248040340287\\
8.06005025125628	0.405201683819713\\
8.13467336683417	0.41627482123627\\
8.20929648241206	0.428760063469517\\
8.28391959798995	0.440942936554396\\
8.35854271356784	0.451965500220125\\
8.43316582914573	0.469440309929817\\
8.50778894472362	0.47598417613058\\
8.58241206030151	0.491063733041133\\
8.6570351758794	0.503629688084223\\
8.73165829145729	0.513670154940323\\
8.80628140703518	0.530422094466779\\
8.88090452261307	0.543985923678448\\
8.95552763819095	0.555958502746328\\
9.03015075376884	0.574995195176389\\
9.10477386934673	0.584221414455404\\
9.17939698492462	0.604488201449413\\
9.25402010050251	0.613257356368993\\
9.3286432160804	0.63422391515378\\
9.40326633165829	0.638959560253457\\
9.47788944723618	0.661003459376505\\
9.55251256281407	0.672592678637504\\
9.62713567839196	0.6959417208706\\
9.70175879396985	0.708596471292016\\
9.77638190954774	0.718050710604812\\
9.85100502512563	0.742248796238273\\
9.92562814070352	0.756670619702346\\
10.0002512562814	0.771092235014353\\
10.0748743718593	0.792626182525518\\
10.1494974874372	0.816452886582108\\
10.2241206030151	0.830257878671731\\
10.298743718593	0.842343139954701\\
10.3733668341709	0.866600711864578\\
10.4479899497487	0.876592184782742\\
10.5226130653266	0.907874243004911\\
10.5972361809045	0.920981176816246\\
10.6718592964824	0.941789236917046\\
10.7464824120603	0.958844203394958\\
10.8211055276382	0.978893659599575\\
10.8957286432161	0.998761975046171\\
10.970351758794	1.02607805250884\\
11.0449748743719	1.04220863119715\\
11.1195979899497	1.06813622340362\\
11.1942211055276	1.08984328242466\\
11.2688442211055	1.10138221584484\\
11.3434673366834	1.13157512102378\\
11.4180904522613	1.15321578976952\\
11.4927135678392	1.18235952555366\\
11.5673366834171	1.20243307078033\\
11.641959798995	1.22066394074797\\
11.7165829145729	1.25316726896221\\
11.7912060301508	1.26722470847296\\
11.8658291457286	1.29169268112636\\
11.9404522613065	1.30895109345296\\
12.0150753768844	1.33338658482894\\
12.0896984924623	1.36633910217999\\
12.1643216080402	1.40051521756916\\
12.2389447236181	1.42558918530796\\
12.313567839196	1.44439753544013\\
12.3881909547739	1.47739677976193\\
12.4628140703518	1.49878229923946\\
12.5374371859296	1.51663229626843\\
12.6120603015075	1.55355921909647\\
12.6866834170854	1.57710487067226\\
12.7613065326633	1.61654213478209\\
12.8359296482412	1.63709959412538\\
12.9105527638191	1.67120832295402\\
12.985175879397	1.68876963727278\\
13.0597989949749	1.71677991851514\\
13.1344221105528	1.76377088550059\\
13.2090452261307	1.78940100028439\\
13.2836683417085	1.81861245629083\\
13.3582914572864	1.8477816245109\\
13.4329145728643	1.87476712639349\\
13.5075376884422	1.90073715434494\\
13.5821608040201	1.95097443083386\\
13.656783919598	1.96627998160562\\
13.7314070351759	1.99972753378414\\
13.8060301507538	2.02855909787324\\
13.8806532663317	2.06306944395287\\
13.9552763819095	2.12627737434575\\
14.0298994974874	2.14275470291698\\
14.1045226130653	2.17018967216683\\
14.1791457286432	2.21221505463514\\
14.2537688442211	2.27102634197462\\
14.328391959799	2.27563977124847\\
14.4030150753769	2.30440858721242\\
14.4776381909548	2.32932128202798\\
14.5522613065327	2.38445503029307\\
14.6268844221106	2.42287891566942\\
14.7015075376884	2.46605849508861\\
14.7761306532663	2.51477504114036\\
14.8507537688442	2.54560020416209\\
14.9253768844221	2.57469917846897\\
15	2.59506622254323\\
};
\addlegendentry{\small{PEB randomized}}

\end{axis}

%\begin{axis}[%
%width=7.778in,
%height=5.833in,
%at={(0in,0in)},
%scale only axis,
%xmin=0,
%xmax=1,
%ymin=0,
%ymax=1,
%axis line style={draw=none},
%ticks=none,
%axis x line*=bottom,
%axis y line*=left,
%legend style={legend cell align=left, align=left, %draw=white!15!black}
%]
%\end{axis}
\end{tikzpicture}%